\documentclass[aps,preprint,prd,showpacs,showkeys,nofootinbib,tightenlines]{revtex4-1}

\usepackage{graphicx}  
\usepackage{amsmath}
\usepackage{bm}  
\usepackage{ulem}
\usepackage{amssymb}
\usepackage{comment}
\usepackage{lineno}
\newcommand{\bea}{\begin{eqnarray}}
\newcommand{\eea}{\end{eqnarray}}
\newcommand{\beq}{\begin{equation}}
\newcommand{\eeq}{\end{equation}}
\newcommand{\bqa}{\begin{eqnarray}}
\newcommand{\eqa}{\end{eqnarray}}

\begin{document}

\title{
Production of $\bm{X(3872)}$ and a Photon  \\ 
in $\bm{e^+e^-}$ Annihilation
}

\author{Eric Braaten}
\email{braaten.1@osu.edu}
\affiliation{Department of Physics,
         The Ohio State University, Columbus, OH\ 43210, USA}

\author{Li-Ping He}
\email{he.1011@buckeyemail.osu.edu}
\affiliation{Department of Physics,
         The Ohio State University, Columbus, OH\ 43210, USA}

\author{Kevin Ingles}
\email{ingles.27@buckeyemail.osu.edu}
\affiliation{Department of Physics,
         The Ohio State University, Columbus, OH\ 43210, USA}

\date{\today}
%\date{November 2007}

\begin{abstract}
If the $X(3872)$ is a weakly bound charm-meson molecule, 
it can be produced in $e^+ e^-$ annihilation by the creation of $D^{*0} \bar D^{*0}$  from a virtual photon
followed by the  rescattering of the P-wave charm-meson pair into the $X$ and a photon.
A triangle singularity produces a narrow peak in the cross section for $e^+ e^- \to X \gamma$ 
2.2~MeV above the $D^{*0} \bar{D}^{*0}$ threshold. 
We predict the normalized cross section in the region of the peak.
We show that the absorptive contribution to the cross section for $e^+ e^- \to D^{*0} \bar D^{*0} \to X \gamma$,
which was calculated previously by Dubynskiy and Voloshin, does not give a good approximation 
to the peak from the triangle singularity.
\end{abstract}

\smallskip
\pacs{14.80.Va, 67.85.Bc, 31.15.bt}
\keywords{
Exotic hadrons, charm mesons, effective field theory.}
\maketitle
%%%%%%%%%%%%%%%%%%%%%%%%%%%%%%%%%%%%%%%%%%
%%%%%%%%%%%%%%%%%%%%%%%%%%%%%%%%%%%%%%%%%%

\section{Introduction}
\label{sec:Introduction}
%%%%%%%%%%%%%%%%%%%%%%%%%%%%%%%%%%%%%%%%%

Since early in this century, a large number of exotic hadrons whose constituents include a heavy quark and its antiquark
have been discovered in high energy physics experiments 
\cite{Chen:2016qju,Hosaka:2016pey,Lebed:2016hpi,Esposito:2016noz,Guo:2017jvc,Ali:2017jda,Olsen:2017bmm,Karliner:2017qhf,Yuan:2018inv,Brambilla:2019esw}. 
The first of these exotic heavy hadrons to be discovered was the $X(3872)$ meson.
It was discovered in 2003 in exclusive decays of $B^\pm$ mesons into $K^\pm X$ 
by observing the decay of $X$ into $J/\psi\, \pi^+\pi^-$ \cite{Choi:2003ue}. 
The $J^{PC}$ quantum numbers of $X$ were eventually determined to be $1^{++}$ \cite{Aaij:2013zoa}.
Its mass  is extremely close to the $D^{*0} \bar D^0$  threshold,
with the difference being only $0.01 \pm 0.18$~MeV \cite{Tanabashi:2018oca}.
This suggests that $X$ is a weakly bound S-wave charm-meson molecule
with the flavor structure
%===============
\begin{equation}
\big| X(3872) \rangle = \frac{1}{\sqrt2}
\Big( \big| D^{*0} \bar D^0 \big\rangle +  \big| D^0 \bar D^{*0}  \big\rangle \Big).
\label{Xflavor}
\end{equation}
%===============
However, there  are alternative models for the $X$
\cite{Chen:2016qju,Hosaka:2016pey,Lebed:2016hpi,Esposito:2016noz,Guo:2017jvc,Ali:2017jda,Olsen:2017bmm,Karliner:2017qhf,Yuan:2018inv,Brambilla:2019esw}.
The $X$ has been observed in 7 different decay modes, many more than any of the other exotic heavy hadrons.
Despite these many decay modes, a consensus on the nature of $X$ has not been achieved.

There may be aspects of the production of $X$ that are more effective at discriminating
between models than the decays of $X$.
If the $X$ is a weakly bound charm-meson molecule, it
can be produced by any reaction that can create its constituents $D^{*0} \bar D^0$ and $D^0 \bar D^{*0}$. 
It can be produced by the creation of $D^{*0} \bar D^0$ and $D^0 \bar D^{*0}$ at short distances 
of order  $1/m_\pi$, where $m_\pi$ is the pion mass,
followed by the binding of the charm mesons into $X$ at longer distances.
The $X$ can also be produced by the  creation of $D^* \bar{D}^*$ at short distances 
followed by the rescattering of the charm-meson pair into $X$ and a pion 
at longer distances \cite{Braaten:2019yua,Braaten:2019sxh}.

One way in which the nature of a hadron can be revealed in its production is through {\it triangle singularities}.
Triangle singularities are kinematic singularities that can arise if three virtual particles that form a triangle 
in a Feynman diagram can all be on their mass shells simultaneously.
There have been several previous investigations of the effects of triangle singularities on the production of
exotic heavy mesons \cite{Szczepaniak:2015eza,Liu:2015taa,Szczepaniak:2015hya,Guo:2017wzr}.
Guo has recently pointed out that  any high-energy process 
that can create $D^{*0} \bar{D}^{*0}$ at short distances in an S-wave channel
will produce $X \gamma$ with a narrow peak near the $D^{*0} \bar{D}^{*0}$ threshold 
due to a charm-meson triangle singularity \cite{Guo:2019qcn}. 
One such process is electron-positron annihilation,
which can create an S-wave $D^{*0} \bar{D}^{*0}$ pair recoiling against a $\pi^0$.
Guo suggested that the peak in the   line shape for $X  \gamma$ due to the  triangle singularity
could be used to  determine the binding energy  of $X$ more accurately than a direct mass measurement.
Because of the charm-meson triangle singularity,
a high-energy process that can create an S-wave $D^* \bar{D}^*$ pair at short distances 
can also produce $X \pi$ with a narrow peak near the $D^* \bar{D}^*$ threshold. 
We noted previously the existence of a such a narrow peak  in the production of $X \pi$ 
in hadron colliders \cite{Braaten:2019sxh}
and in $B$ meson decays into $K X \pi$ \cite{Braaten:2019yua}, 
but we did not recognize the connection to triangle singularities.

The quantum numbers $1^{++}$ of the $X$ imply that $X\gamma$ can be produced by 
$e^+ e^-$ annihilation into a virtual photon.  The virtual photon can create
$D^{*0} \bar{D}^{*0}$ at short distances in a P-wave channel, and the charm-meson pair
can subequently rescatter into $X \gamma$.
The production of $X \gamma$ in $e^+ e^-$ annihilation near the $D^{*0} \bar D^{*0}$ threshold
was first discussed by Dubynskiy and Voloshin  \cite{Dubynskiy:2006cj}.
They calculated the absorptive contribution to the cross section from 
$e^+ e^-$ annihilation into on-shell charm mesons $D^{*0} \bar{D}^{*0}$ 
followed by their rescattering into $X \gamma$.
They predicted that the  cross section has a narrow peak only a few MeV above the $D^{*0} \bar D^{*0}$ threshold.
In retrospect, the narrow peak comes from a triangle singularity.
In Ref.~\cite{Braaten:2019gfj}, we calculated the cross section for  $e^+ e^- \to X \gamma$ in the energy region
near the $D^{*0} \bar{D}^{*0}$ threshold, including the  dispersive  contributions as well as the absorptive contributions.  
The dispersive contributions have a significant effect on the shape of the narrow peak.
We also predicted the normalization of the cross section for  $e^+ e^- \to X \gamma$ 
using a fit to the cross section for  $e^+ e^- \to D^{*+} D^{*-}$ by Uglov {\it et al.}\ \cite{Uglov:2016orr}.

The production of $X \gamma$ in $e^+ e^-$ annihilation has been studied by the BESIII collaboration 
\cite{Ablikim:2013dyn,Ablikim:2019zio}.
The cross section was measured at the center-of-mass energies ranging from 4.008~GeV to 4.6~GeV.
The cross section was not measured at energies near the $D^{*0} \bar D^{*0}$ threshold at 4.014~GeV,
which is where the narrow peak from the charm-meson triangle singularity is predicted to appear.
The observation of this peak would provide strong evidence 
in support of the identification of the $X$ as a charm-meson molecule.

In this paper, we describe in detail the calculation of the cross section for $e^+e^-\to X \gamma$. 
In Section~\ref{sec:D*D*}, we calculate the cross section for  $e^+e^- \to D^{*0} \bar D^{*0}$ near the threshold. 
We determine the normalization of the cross section by using a previous fit to Belle data on
$e^+e^- \to D^{*+} D^{*-}$.
In Section~\ref{sec:Xgamma-low}, we calculate the cross section for  $e^+e^-\to X \gamma$ from the creation 
of a P-wave $D^{*0} \bar D^{*0}$ pair by a virtual photon followed by the rescattering of the charm mesons 
to $X\gamma$. We reduce  the amplitude to a scalar loop integral. 
In Section~\ref{sec:Triangle}, we calculate the loop amplitude analytically and show that there is a triangle singularity.
We predict the  normalized cross section  for $e^+e^-\to X \gamma$ at energies 
in the region near the  peak from the triangle singularity.  In Section~\ref{sec:Wavefunction}, we 
express the loop amplitude in terms of the Schr\"odinger wavefunction for the bound state.
We point out the differences from the wavefunction used by Dubynskiy and Voloshin  in Ref.~\cite{Dubynskiy:2006cj}.
In Section~\ref{sec:Absorptive}, we calculate the absorptive contribution to the cross section 
for $e^+e^-\to X \gamma$ from intermediate charm mesons $D^{*0} \bar D^{*0}$ that are on their mass shells.
We show that it does not provide a good approximation to the peak in the cross section from the triangle singularity.
Our results are summarized in Section~\ref{sec:Summary}.
In an Appendix, we present a diagrammatic derivation of the Schr\"odinger wavefunction of the $X$ 
in a frame where its momentum is nonzero.
We also present a prescription for  calculating the cross section for producing the 
$X$ resonance feature in cases where the $X$ is not a narrow bound state.

%\newpage

%%%%%%%%%%%%%%%%%%%%%%%%%%%%%%%%
\section{Production of   $\bm{D^{*0} \bar{D}^{*0}}$ near  threshold}
\label{sec:D*D*}
%%%%%%%%%%%%%%%%%%%%%%%%%%%%%%%

%%%%%%%%%%%%%%%%%%%%%%%%%%%%%%%%%%%%%%%%%%%%%%%%
\begin{figure}[ht]
\includegraphics*[width=0.45\linewidth]{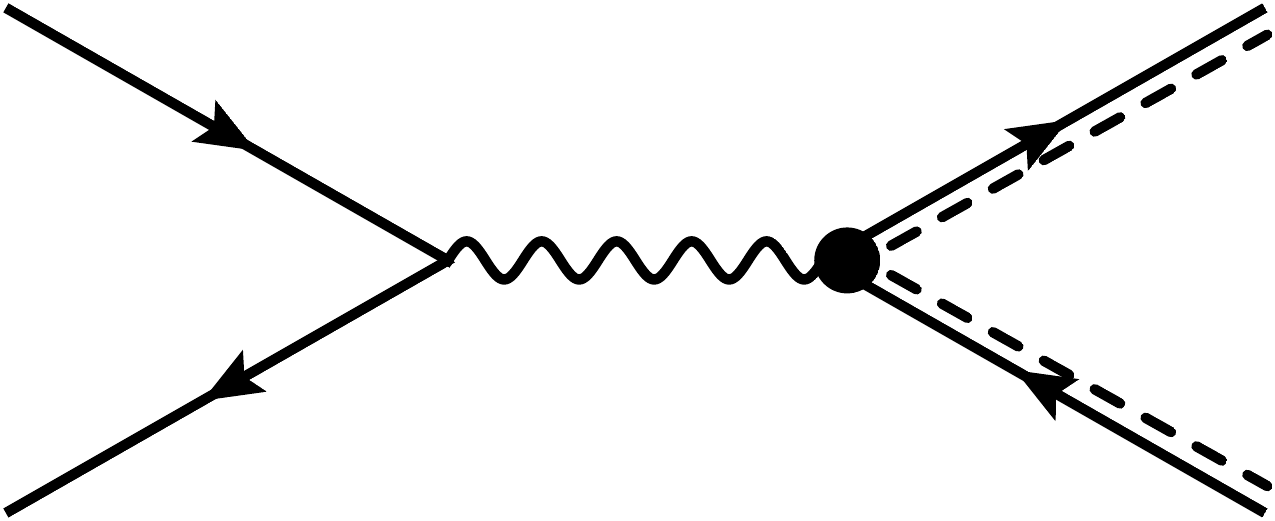} 
\caption{Feynman diagram for $e^+e^-\to  D^{*0} \bar D^{*0}$. 
The spin-1 charm mesons $D^{*0}$ and $\bar D^{*0}$ are represented by double lines 
consisting of a dashed line and a solid line with an arrow.
}
\label{fig:eetoD*D*}
\end{figure}
%%%%%%%%%%%%%%%%%%%%%%%%%%%%%%%%%%%%%%%%%%%%%%

A pair of spin-1 charm mesons $D^{*0} \bar D^{*0}$ can be produced from the annihilation of $e^+e^-$ 
into a virtual photon. The Feynman diagram for this process is shown in Fig.~\ref{fig:eetoD*D*}.
We use nonrelativistic normalizations for the  charm mesons in the final state.
In the center-of-momentum (CM)  frame, the matrix element has the form
%===============
\begin{equation}
\mathcal{M}= -i\frac{e^2}{s}\, \bar v \gamma^i u\,J^i ,
\label{MeetoD*D*}
\end{equation}
%===============
where $\sqrt{s}$ is the invariant mass, $\bar v$ and $u$ are the spinors for the colliding
$e^+$ and $e^-$, and $\bm{J}$ is the matrix element of the electromagnetic current
between the QCD vacuum and the $D^{*0} \bar D^{*0}$ state.
Near the threshold for producing $D^{*0} \bar D^{*0}$, the charm-meson pair is produced in a P-wave state 
with total spin 0 or 2.  The matrix element of the electromagnetic current that creates  $D^{*0}$ and $\bar D^{*0}$ 
with momenta $+\bm{k}$ and $-\bm{k}$ and with polarization vectors $\bm{\varepsilon}$ and $\bar{\bm{\varepsilon}}$  
can be expressed as 
$J^i=\mathcal{A}^{ijkl} k^j\varepsilon^{*k}\bar{\varepsilon}^{*l}$.
The Cartesian tensor $\mathcal{A}^{ijkl}$ is
%===============
\begin{equation}
\mathcal{A}^{ijkl}= A_0 \, \delta^{ij} \delta^{kl} 
+ \frac{3}{2\sqrt{5}} A_2 \left(\delta^{ik} \delta^{jl}+\delta^{il} \delta^{jk} - \frac23 \delta^{ij} \delta^{kl}  \right) ,
\label{A[eetoD*D*]}
\end{equation}
%===============
where  $A_0$ and $A_2$ are amplitudes for creating $D^{*0} \bar D^{*0}$ with total spin 0 and 2, respectively.  
The numerical prefactor of $A_2$ in Eq.~\eqref{A[eetoD*D*]} was chosen for later convenience.

The  differential cross section for producing  $D^{*0} \bar D^{*0}$ with scattering angle $\theta$ is 
%===============
\begin{equation}
\frac{d\sigma}{d\Omega}= \frac{3 \alpha^2 M_{*0}}{2s^2} k^3
\left[ |A_0|^2 (1 - \cos^2\theta) + |A_2|^2 \frac{7 - \cos^2\theta}{10} \right]   ,
\label{dsigma/dOmega}
\end{equation}
%===============
where  $M_{*0}$ is the mass of the  $D^{*0}$
and $k$ is the relative momentum of the $D^{*0} \bar D^{*0}$ pair:  $k =[ M_{*0} (\sqrt{s} - 2 M_{*0} )]^{1/2}$. 
The cross section for $e^+e^-$ annihilation into $D^{*0} \bar D^{*0}$ near the  threshold is
%===============
\begin{equation}
\sigma[ e^+ e^- \to D^{*0} \bar D^{*0}] = \frac{4\pi \alpha^2 M_{*0}}{s^2}
\left[ |A_0|^2  + |A_2|^2  \right] k^3  .
\label{sigma}
\end{equation}
%===============
The absolute values of the two amplitudes $A_0$ and $A_2$ could in principle be determined experimentally
from the value of the cross section and from the angular distribution at a single energy near the threshold.

%%%%%%%%%%%%%%%%%%%%%%%%%%%%%%%%%%%%%%%%%%%%%%%%
\begin{figure}[t]
\includegraphics*[width=0.7\linewidth]{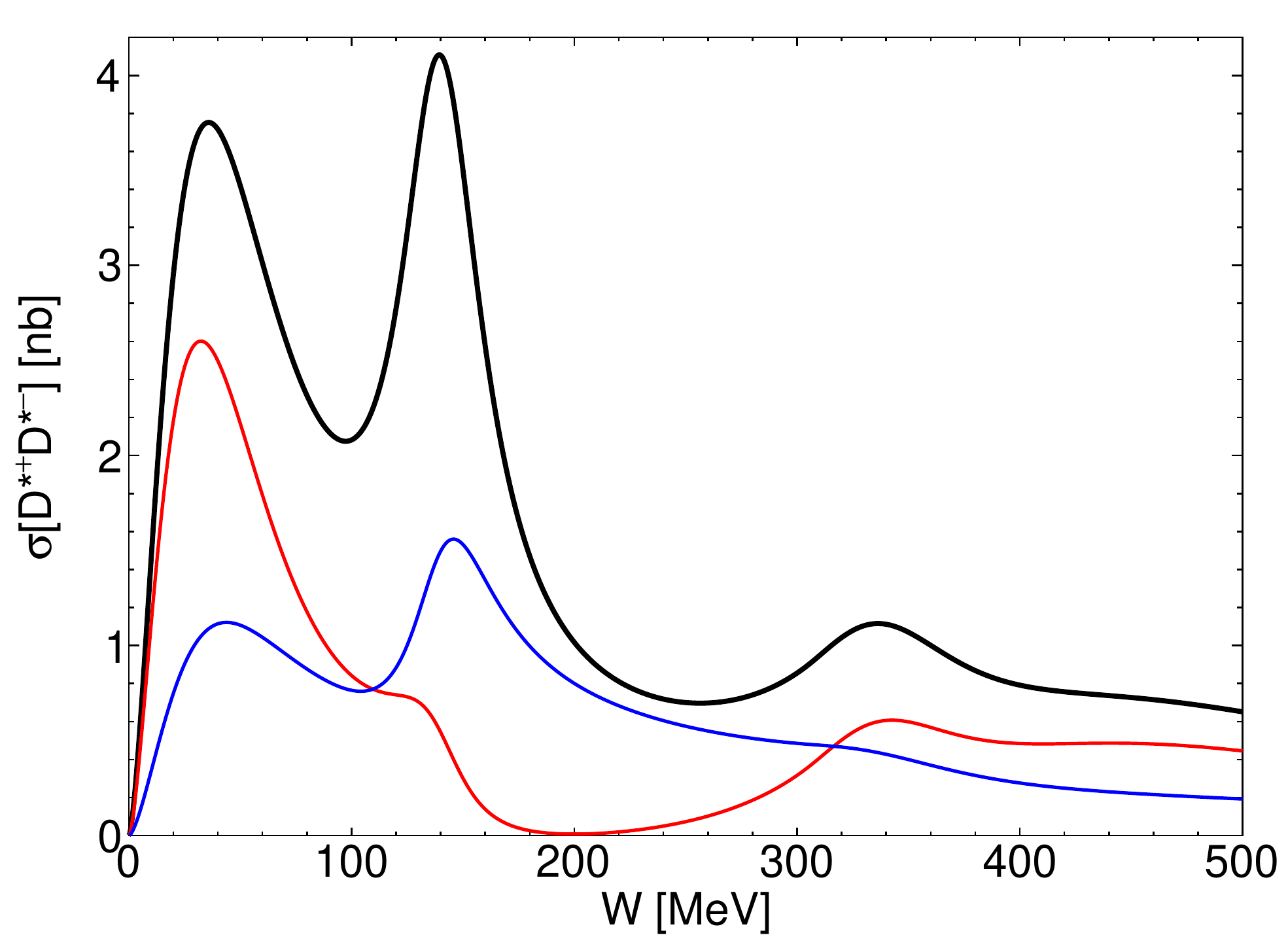} 
\caption{
Cross section for $e^+ e^- \to D^{*+} D^{*-}$ as a function of the center-of-mass energy 
$W$ relative to the $D^{*+} D^{*-}$ threshold.
The curves are fits to Belle data  by Uglov {\it et al.}\ \cite{Uglov:2016orr}: full cross section (thicker black curve), 
spin-2 P-wave contribution (taller thin red curve), and spin-0 P-wave contribution (shorter thin blue curve).
}
 \label{fig:sigma-D*+D*-}
%\vspace*{0.0cm}
\end{figure}
%%%%%%%%%%%%%%%%%%%%%%%%%%%%%%%%%%%%%%%%%%%%%%

The Belle collaboration has measured exclusive cross sections for $e^+ e^-$ annihilation 
into several pairs of charm mesons, including $D^{*+} D^{*-}$  \cite{Abe:2006fj,Pakhlova:2008zza}. 
Uglov {\it et al.}\  have analyzed the Belle data  using a unitary approach 
based on a coupled channel model  \cite{Uglov:2016orr}. 
They included a  spin-2 F-wave amplitude for $e^+ e^- \to D^* \bar D^*$ as well as 
spin-0 and spin-2 P-wave amplitudes.  Their fit to the cross section for $e^+ e^- \to D^{*+} D^{*-}$  
as a function of the center-of-mass energy $W$ relative to the $D^{*+} D^{*-}$ threshold
is shown in Fig.~\ref{fig:sigma-D*+D*-},
along with  the spin-0 and spin-2 P-wave contributions.\footnote{
The curves in Figure~1(c) of Ref.~\cite{Uglov:2016orr} are mislabeled in the figure caption.}
The fitted cross section increases to a local maximum of 3.8~nb at $W=36$~MeV,  
and then decreases to a local minimum of 2.1~nb at $W=97$~MeV. 
At the energy of the first local minimum, the spin-2 F-wave contribution has increased to 21\% of the cross section. 
Thus the P-wave contributions alone give a good approximation to the cross section 
for $W$ up to about 100~MeV.

\begin{figure}[t]
\includegraphics*[width=0.7\linewidth]{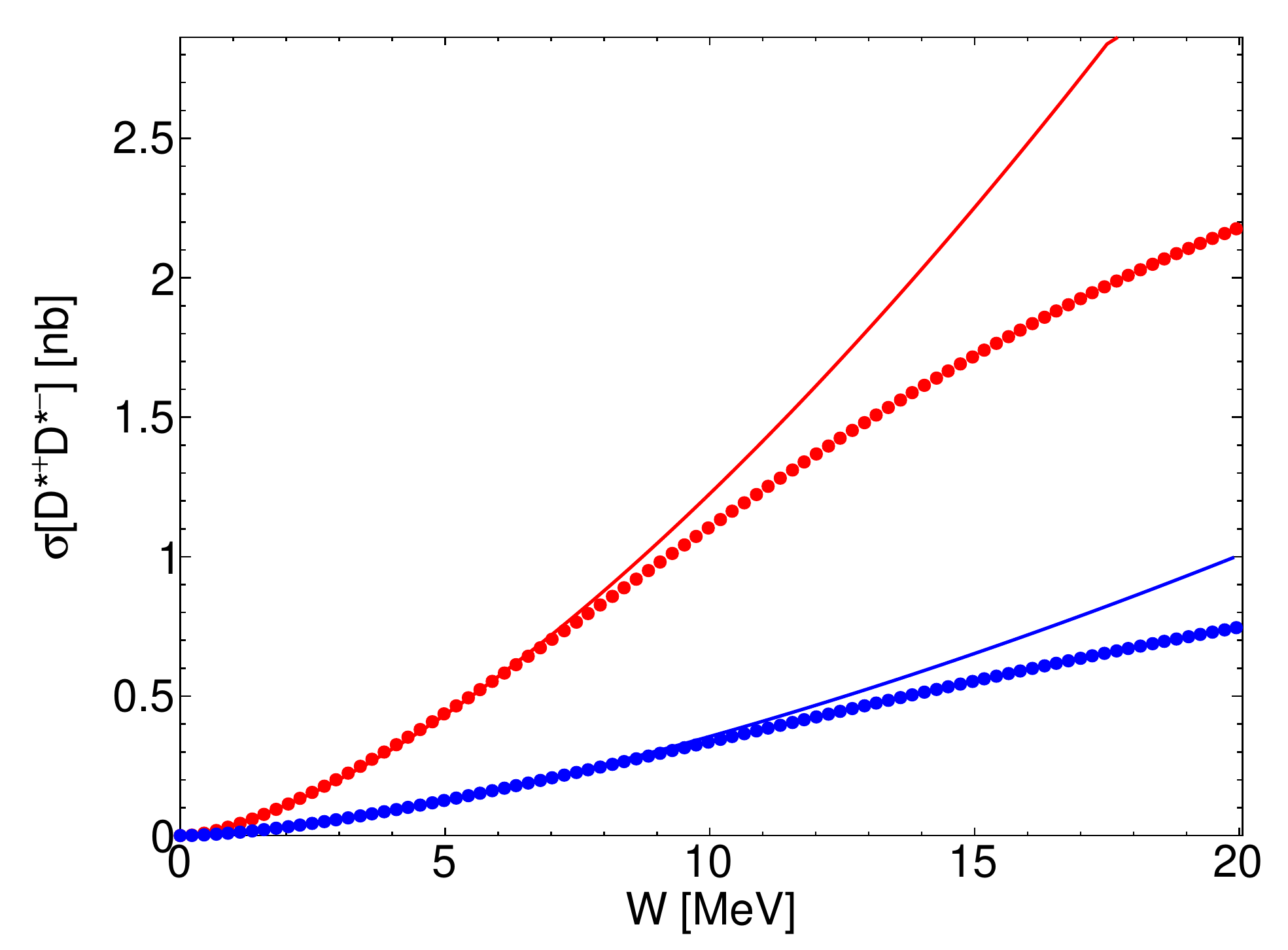} 
\caption{
Cross section for $e^+ e^- \to D^{*+} D^{*-}$ near threshold as a function of the center-of-mass energy 
$W$ relative to the $D^{*+} D^{*-}$ threshold.
The dots are the spin-2 P-wave contribution (higher red dots) and the spin-0 P-wave contribution (lower  blue dots) 
from  Uglov {\it et al.}\ \cite{Uglov:2016orr}, which are shown as curves in Fig.~\ref{fig:sigma-D*+D*-}.
The curves are fits to the analog of Eq.~\eqref{sigma}, and they determine the coefficients in Eq.~\eqref{A0,A2}.
}
\label{fig:sigma-D*+D*-:threshold}
%\vspace*{0.0cm}
\end{figure}

Near the $D^{*+} D^{*-}$ threshold at 4020.5~MeV, 
the  spin-0 and spin-2 P-wave contributions to the cross sections have the $k^3$ behavior in Eq.~\eqref{sigma}.
A fit to the two terms in the cross section in Eq.~\eqref{sigma}, 
with $M_{*0}$ replaced by the mass $M_{*1}$ of the $D^{*+}$ and $k =[ M_{*1} (\sqrt{s} - 2 M_{*1} )]^{1/2}$, gives
%===============
\begin{equation}
|A_0| = 8~\mathrm{GeV}^{-1}, \qquad |A_2| = 15~\mathrm{GeV}^{-1} .
\label{A0,A2}
\end{equation}
%===============
These coefficients have natural magnitudes of order $1/m_\pi$.
The fits to the  spin-0 and spin-2 P-wave contributions to the cross sections
of Uglov {\it et al.}\  are shown in Fig.~\ref{fig:sigma-D*+D*-:threshold}.
The fits are very good for $W$ up to about 10~MeV.

%%%%%%%%%%%%%%%%%%%%%%%%%%%%%%%%%%%%%%%%%%%%%%%%
%%%%%%%%%%%%%%%%%%%%%%%%%%%%%%%%%%%%%%%%%%%%%%

We wish to relate the amplitudes $A_0$  and $A_2$ for $e^+ e^- \to D^{*0} \bar D^{*0}$ 
to the corresponding amplitudes for $e^+ e^- \to D^{*+} D^{*-}$.
An amplitude $A_i$ for $D^{*+} D^{*-}$ is the sum of an isopin-0 amplitude and an isospin-1 amplitude, 
while the corresponding amplitude for $D^{*0} \bar D^{*0}$  is the difference.
Fig.~\ref{fig:sigma-D*+D*-} shows that the spin-2 P-wave contribution to the cross section for 
$e^+ e^- \to D^{*+} D^{*-}$ has a strong peak near 4050~MeV, and that the spin-0 P-wave contribution
also has a peak near that energy.  This strongly suggests that the amplitudes $A_2$ and $A_0$ 
are dominated by the  $\psi(4040)$ charmonium resonance, which has mass 4039~MeV
and width 80~MeV. Since the width of this isospin-0 resonance is  much larger than the 6.8~MeV
difference between the charm-meson pair thresholds, the amplitudes $A_2$ and $A_0$ 
for $D^{*0} \bar D^{*0}$ must also be dominated by the  $\psi(4040)$ resonance.
We assume the isospin-1 amplitudes are negligible compared to the resonant isospin-0 amplitudes.
We therefore approximate the amplitudes  $A_0$ and $A_2$
for $e^+ e^- \to D^{*0} \bar D^{*0}$ by the corresponding amplitudes for $e^+ e^- \to D^{*+} D^{*-}$.
We can predict the cross section for $e^+ e^- \to D^{*0} \bar D^{*0}$ 
near its threshold at 4013.7~MeV
by inserting the values of $|A_0|$  and $|A_2|$ in Eq.~\eqref{A0,A2} 
into Eq.~\eqref{sigma}.

The Belle data on $e^+ e^-$ annihilation into charm-meson pairs in 
Refs.~\cite{Abe:2006fj} and \cite{Pakhlova:2008zza} 
has also been analyzed by Du, Mei{\ss}ner, and Wang  using an approach that takes into account 
P-wave coupled channel effects by solving Lippmann-Schwinger equations with contact interactions 
between the charm mesons \cite{Du:2016qcr}.
They presented their fit to the cross section for $D^{*+} D^{*-}$ in the form of histograms with 20~MeV bins.
It is therefore not possible to determine $|A_0|$  and $|A_2|$ from their results.
A comparison of their results with those of Ref.~\cite{Uglov:2016orr} 
can however give some indication of the possible size of theoretical errors.
In Ref.~\cite{Du:2016qcr}, the spin-0 and spin-2 P-wave contributions to the cross section were given 
separately only at the single  invariant mass $\sqrt{s} = 4.040$~GeV,
which is about 17~MeV above the $D^{*+} D^{*-}$ threshold.
The cross sections are 1.52~nb and 1.23~nb, respectively,  
compared to 0.75~nb and 2.19~nb   from the fit in Ref.~\cite{Uglov:2016orr}.
The sum of the two cross sections in Ref.~\cite{Du:2016qcr} is only about 6\% smaller 
than their sum  from the fit in   Ref.~\cite{Uglov:2016orr}.
However the ratio 0.81 of the spin-2 and spin-0 cross sections in  Ref.~\cite{Du:2016qcr} 
is  significantly smaller   than the ratio  2.92 from the fit in Ref.~\cite{Uglov:2016orr}. 
We conclude  that $ |A_0|^2  + |A_2|^2$ can be determined more accurately 
by fitting cross sections than $ |A_2|/|A_0|$.
The phases of the amplitudes $A_0$ and $A_2$
can be chosen so that a linear combination of the amplitudes has no interference.
For this choice, the values in Eq.~\eqref{A0,A2} can be expressed as
%===============
\begin{equation}
 |A_0|^2  + |A_2|^2 = 280~\mathrm{GeV}^{-2}, \qquad 
 A_2/A_0 = \pm 1.9 \, i.
\label{A^2,A2/A0}
\end{equation}
%===============
We will take these to be our preferred values for the amplitudes.
However we will consider all possible complex values for the ratio $A_2/A_0$
that are consistent with the value of $|A_0|^2  + |A_2|^2$ in Eq.~\eqref{A^2,A2/A0}.

%\newpage

%%%%%%%%%%%%%%%%%%%%%%%%%%%%%%%%
\section{Production of   $\bm{X+ \gamma}$ near  the $\bm{D^{*} \bar{D}^{*}}$ threshold}
\label{sec:Xgamma-low}
%%%%%%%%%%%%%%%%%%%%%%%%%%%%%%%

If the $X(3872)$ is a weakly bound charm-meson molecule, its constituents are
the superposition of charm mesons in Eq.~\eqref{Xflavor}.
The reduced mass of $D^{*0} \bar D^0$ is $\mu=M_{*0}M_0/(M_{*0}\!+\!M_0)$,
where $M_0$ is the mass of the $D^0$.
The mass difference between the $D^{*0}$ and $D^0$ is $\delta = M_{*0}\!-\!M_0= 142.0$~MeV.
The decay  width of the $D^{*0}$ can be predicted from measurements of $D^*$ decays: 
$\Gamma_{*0} = (55.9 \pm 1.6)$~keV  \cite{Rosner:2013sha}. 
The present value of the difference $E_X$ between the mass of the $X$ 
and the energy of the $D^{*0} \bar D^0$ scattering threshold is \cite{Tanabashi:2018oca}
%===============
\begin{equation}
E_X \equiv M_X - (M_{*0}\!+\!M_0) =( +0.01 \pm 0.18)~\mathrm{MeV}.
\label{EX-exp}
\end{equation}
%===============
The central value corresponds to a charm-meson pair just above the scattering threshold.
The value lower by $1\sigma$ corresponds to a bound state with binding energy $|E_X| =0.17$~MeV.  

%%%%%%%%%%%%%%%%%%%%%%%%%%%%%%%%%%%%%%%%%%%%%%%%
\begin{figure}[ht]
\includegraphics*[width=0.45\linewidth]{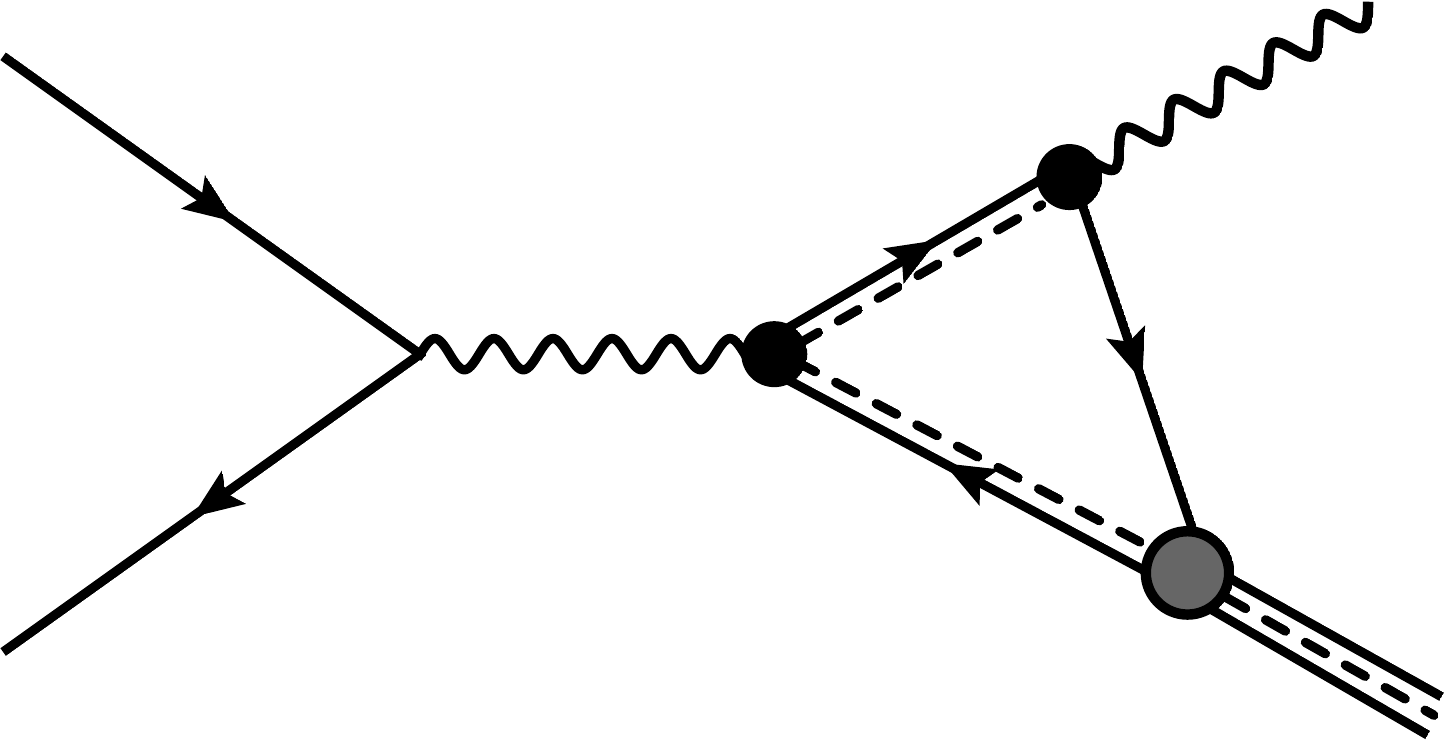} \quad
\includegraphics*[width=0.45\linewidth]{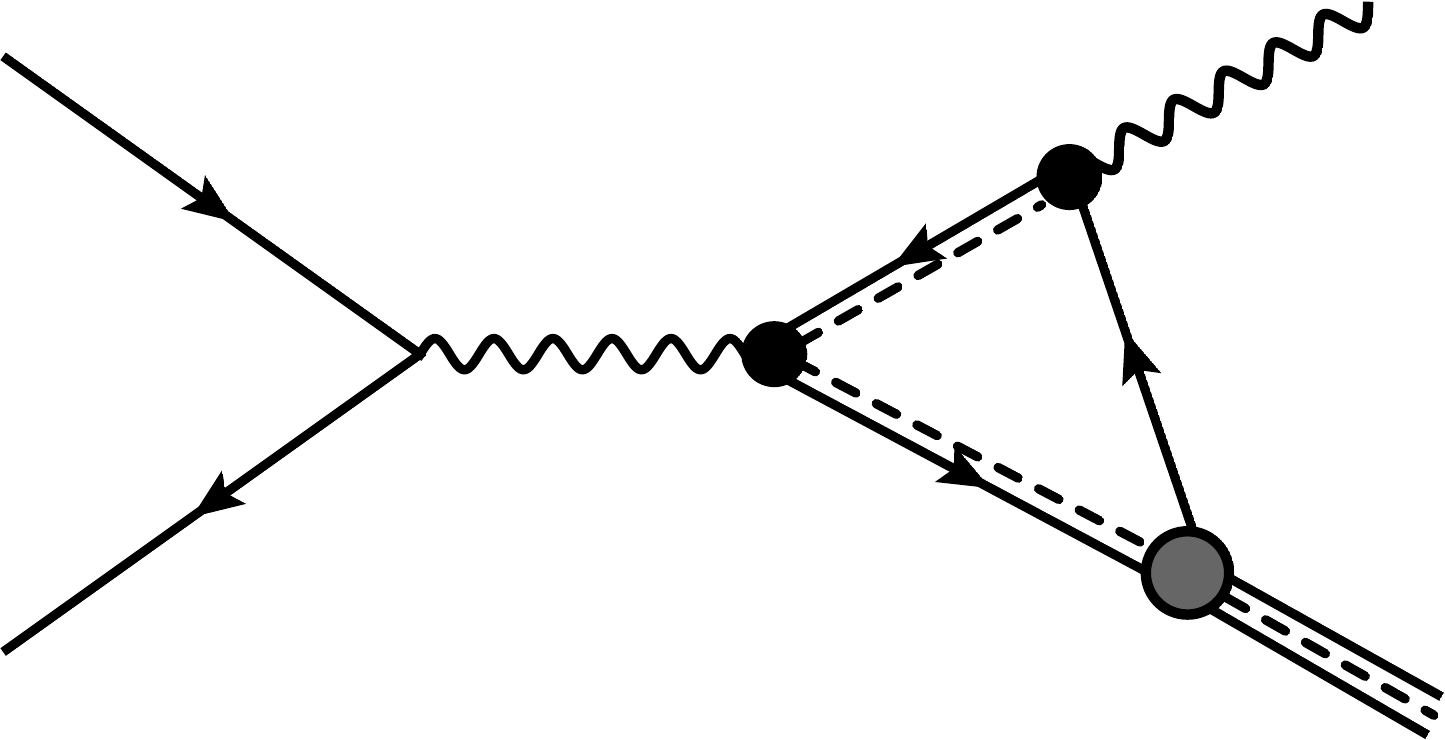} 
\caption{Feynman diagrams for $e^+e^-\to \gamma X$ from rescattering of  $D^{*0} \bar D^{*0}$. 
The $X$ is represented by a triple line consisting of two solid lines and a dashed line. 
The spin-0 charm mesons $D^0$ and $\bar D^0$ are represented by solid lines with an arrow.
}
\label{fig:eetogammaX}
\end{figure}
%%%%%%%%%%%%%%%%%%%%%%%%%%%%%%%%%%%%%%%%%%%%%%

The $X$ can be produced in $e^+e^-$ annihilation through the creation of $D^{*0} \bar D^{*0}$
by a virtual photon followed by the rescattering of the charm-meson pair into $X\gamma$.
The Feynman diagrams for this process are shown in Fig.~\ref{fig:eetogammaX}.
The vertex for the virtual photon with vector index $i$ to create $D^{*0}$ and $\bar D^{*0}$ 
with momenta $+\bm{k}$ and $-\bm{k}$ and with vector indices $m$ and $n$ is $e \mathcal{A}^{ijmn} k^j$,
where the Cartesian tensor is given in Eq.~\eqref{A[eetoD*D*]}.
The vertex for the transition of $D^{*0}$ to $D^0\gamma$ with a photon of momentum  $\bm k$
is $-e\nu \epsilon^{ijm} k^m$, where $i$ and $j$ are the vector indices of $D^{*0}$ and $\gamma$.  
The transition magnetic moment $e\nu$ can be determined from the radiative decay width of $D^{*0}$:
%===============
\begin{equation}
\Gamma[D^{*0} \to D^0 \gamma] =  \frac{4\alpha \nu^2 \omega^3}{3(1+\omega/M_0)},
\label{GammaD*0}
\end{equation}
%===============
where the photon energy $\omega$ satisfies $\omega+\omega^2/2M_0= \delta$.
The radiative width of the $D^{*0}$ can be predicted from measurements of $D^*$ decays: 
$\Gamma[D^{*0} \to D^0 \gamma] = (21.2 \pm 1.6)$~keV  \cite{Rosner:2013sha}. 
This determines the transition magnetic moment:  $\nu= 0.95~\mathrm{GeV}^{-1}$.
The binding of $D^{*0} \bar D^{0}$ or $D^{0} \bar D^{*0}$ into $X$ 
can be described within an effective field theory called XEFT \cite{Fleming:2007rp,Braaten:2015tga}.
The vertices for the couplings of $D^{*0}\bar D^0$ to $X$ and $D^{0}\bar D^{*0}$ to $X$ can be expressed as 
%===============
\begin{equation}
i (\pi \gamma_X/\mu^2)^{1/2}\, \delta^{kl},
\label{vertexX}
\end{equation}
%===============
where $\gamma_X$ is a parameter with dimensions of momentum
and $k$ and $l$ are the vector indices of the  spin-1 charm meson and the $X$ \cite{Braaten:2010mg}.
In the Appendix, this vertex is derived from the residue of the pole in the elastic scattering amplitude 
for the charm-meson pair. 

We will assume in the body of this paper that the $X$ is a narrow bound state whose binding energy  $|E_X|$
is significantly larger than the decay width $\Gamma_{*0} \approx 56$~keV of the $D^{*0}$.
The energy of the $X$ can therefore be expressed in terms of the 
positive real binding momentum $\gamma_X$ as
%===========
\begin{eqnarray}
 E_X = -\gamma_X^2/2\mu.
\label{EX-gammaX} 
\end{eqnarray}
%===========
For $E_X = -0.17$~MeV, the binding momentum is $\gamma_X=18$~MeV.
In the simplest plausible model for the resonant scattering amplitude of the charm mesons,
this binding momentum coincides with the parameter $\gamma_X$ in the vertex in Eq.~\eqref{vertexX}.
We will therefore use Eq.~\eqref{EX-gammaX} to determine the real parameter $\gamma_X$
from the binding energy  $|E_X|$.
The production of the $X$ resonance feature in the more general case 
where $X$ is not a narrow bound state is considered in the Appendix.

The matrix element for $e^+e^-\to X \gamma$ is the sum of the two diagrams  in Fig.~\ref{fig:eetogammaX}.  
We use nonrelativistic propagators for the charm mesons.
The matrix element for producing $X$ and $\gamma$ with momenta $\bm q$ and $-\bm q$
and with polarization vectors $\bm \varepsilon_X$ and $\bm \varepsilon_\gamma$ can be expressed as
%===========
\begin{equation}
\mathcal{M} = \frac{2e^3 \nu M_{*0}M_0}{s\, \mu} \, \bar v \gamma^i u \, \mathcal{J}^i\, F(W),
\label{MeetoXgamma}
\end{equation}
%===========
where $F(W)$  is a function of the center-of-mass energy $W = \sqrt{s}-2M_{*0}$ 
relative to the $D^{*0}\bar D^{*0}$ threshold.
We will refer to  $F(W)$ as the  {\it loop amplitude}, because it can be expressed as 
an integral over the undetermined energy and momentum in the charm-meson loops in Fig.~\ref{fig:eetogammaX}:
%===========
\begin{eqnarray}
 F(W)&=& 
 i  \frac{\sqrt{\pi \gamma_X} }{M_{*0}M_0}
    \int\!\! \frac{d^3k}{(2\pi)^3}\, \bm q \cdot \bm k \! \int\!\! \frac{d\omega}{2\pi} 
\, \frac{1}{\omega - \bm k^2/(2M_{*0})+ i \Gamma_{*0}/2}
\nonumber\\
&& \hspace{0cm} \times
   \frac{1}{W -\omega - \bm k^2/(2M_{*0})+ i\Gamma_{*0}/2}\,
   \frac{1}{W - (|\bm q| - \delta)  - \omega - (\bm q -\bm k)^2/(2M_0) + i \epsilon}.
\label{Fd4k} 
\end{eqnarray}
%===========
To obtain the scalar loop integral in Eq.~\eqref{Fd4k},  we used
rotational symmetry to replace a factor of $k^i$ inside the momentum integral
by  $(\bm q \cdot \bm k)q^i/\bm q^2$.  The resulting expression for the current $\mathcal{J}^i$ is
%===========
\begin{equation}
\bm{\mathcal{J}}= 
\left(A_0 - \frac{1}{\sqrt{5}} A_2 \right)  (\hat{\bm q} \times \bm \epsilon_\gamma \cdot \bm \epsilon_X)\, \hat{\bm q}
+\frac{3}{2\sqrt{5}} A_2 (\hat{\bm q} \cdot \bm \epsilon_X)  \hat{\bm q} \times \bm \epsilon_\gamma .
\label{currentJ}
\end{equation}
%===========

In the production of $X\gamma$ invariant mass $\sqrt{s}= 2M_{*0} + W$,
the center-of-mass energy $W$ relative to the $D^{*0}\bar D^{*0}$ threshold is determined by energy conservation:
%==============
\begin{equation}
W = (q -\delta)+ \frac{q^2}{2(M_{*0} \!+\! M_0)} + E_X,
\label{Econservation}
\end{equation}
%==============
where $q$ is the photon energy and $\delta = M_{*0}\!-\!M_0$.
We have used the Galilean-invariant approximation for the kinetic energy of the $X$,
in which its kinetic mass is the sum of the masses of $D^{*0}$ and $\bar D^0$.
Assuming $W$ is less than or of order $\delta$, 
we can solve Eq.~\eqref{Econservation} for the photon energy $q$ 
as an expansion in powers of $\delta/(M_{*0}\!+\!M_0)$ and $E_X/\delta$:
%===========
\begin{equation}
q =  ( \delta+W) - \frac{(\delta+W)^2}{2M_X} +\frac{(\delta+W)^3}{2M_X^2} -E_X + \ldots.
\label{q-W}
\end{equation}
%===========

The  differential cross section for producing  $X$ with  scattering angle $\theta$ is
%===============
\begin{equation}
\frac{d\sigma}{d\Omega}=\frac{16 \pi \alpha^3 \nu^2(M_{*0}\!+\!M_0)^2}{s^2[1+q/(M_{*0} \!+\! M_0)]}\,  
q\, \big| F(W) \big|^2
\left[\left| A_0 - \frac{1}{\sqrt{5}} A_2\right|^2 ( 1- \cos^2 \theta ) + \frac{9}{40}|A_2|^2 (1+ \cos^2 \theta)\right] .
\label{dsigma/dOmega:Xgamma}
\end{equation}
%===============
We have used nonrelativistic phase space for $X$ and relativistic phase space for the photon.
The cross section for $e^+ e^-$ annihilation into $X \gamma$ near the $D^{*0}\bar D^{*0}$ threshold is
%==============
\begin{equation}
\sigma[e^+ e^- \to X \gamma]=
\frac{128 \pi^2\alpha^3 \nu^2 (M_{*0}\!+\! M_0)^2}{3 s^2  [1+q/(M_{*0} \!+\! M_0)]} 
\left(\left| A_0 - \frac{1}{\sqrt{5}} A_2\right|^2  + \frac{9}{20} |A_2|^2 \right)\,
q \big| F(W) \big|^2.
\label{sigmagammaXana}
\end{equation}
%==============
The factor that depends on $A_0$  and $A_2$ differs from the value of $|A_0|^2  + |A_2|^2$ in Eq.~\eqref{A^2,A2/A0}
by a multiplicative factor that depends on the value of the ratio $A_2/A_0$.
For our preferred values $A_2/A_0= \pm 1.9\, i$ in Eq.~\eqref{A^2,A2/A0}, 
the multiplicative factor is 0.73. If we allow all possible complex values of $A_2/A_0$
consistent with the value of $|A_0|^2  + |A_2|^2$ in Eq.~\eqref{A^2,A2/A0},
the multiplicative factor can range from 0.34 to 1.31.

%\newpage

%%%%%%%%%%%%%%%%%%%%%%%%%%%%%%%%
\section{Peak from the Triangle Singularity}
\label{sec:Triangle}
%%%%%%%%%%%%%%%%%%%%%%%%%%%%%%%

The loop amplitude $F(W)$  in Eq.~\eqref{Fd4k}  has a kinematic singularity called a {\it triangle singularity} 
\cite{Szczepaniak:2015eza,Liu:2015taa,Szczepaniak:2015hya,Guo:2017wzr} 
in the limit where the binding energy of $X(3872)$ is 0 and the decay width of the $D^{*0}$ is 0. 
The singularity arises from  the integration region where the three charm mesons whose  lines form triangles
in the diagrams in Fig.~\ref{fig:eetogammaX} are all on their mass shells simultaneously. 
The two charm mesons that become constituents of the $X$ are both  on their mass shells 
in the limit where the binding energy is 0.  There is a specific energy $W_\triangle$ where
the spin-1 charm meson that  emits the photon can also be on its mass shell.
If  $\gamma_X =0$  and $\Gamma_{*0} = 0$, $F(W)$ has a logarithmic divergence at $W_\triangle$.
When either $\gamma_X$ or $\Gamma_{*0}$ is nonzero, $F(W)$ has a  narrow peak near $W_\triangle$.

If the integral over the loop energy $\omega$ in Eq.~\eqref{Fd4k} is evaluated by closing the contour in the lower half-plane,
the only pole in $\omega$ is from the propagator of the spin-1 charm meson that becomes a constituent of the $X$.
The denominators of the  propagator of the spin-1 charm meson that emits a photon 
and the propagator of the spin-0 charm meson that becomes a constituent of the $X$ 
can be combined into a single denominator by introducing an integral over a Feynman parameter $x$.
The integral over the loop 3-momentum $\bm{k}$ can  be evaluated analytically.
The resulting loop amplitude can be expressed as
%===========
\begin{equation}
F(W) =  i\frac{\mu \sqrt{\pi \gamma_X}}{4\pi M_0} q \int_0^1 dx \frac{\sqrt{c}\, x}{\sqrt{ a + b x + c x^2 }}.
\label{eetoXgammadx}
\end{equation}
%===========
The coefficients of the polynomial inside the square root are
%============
\begin{subequations}
\begin{eqnarray}
a &=&   k^2+i M_{*0}\Gamma_{*0} ,
    \\
b &=&  -\big[   (\mu/M_0)^2   q^2 +   k^2 + \gamma_X^2\big] - i  (\mu/M_0)M_{*0}\Gamma_{*0} ,
\label{coeffb}
    \\
c &=&  (\mu/M_0)^2  q^2 ,
\end{eqnarray}
\label{coefficientsabc}%
\end{subequations}
%===========
where $k^2=M_{*0}W$, $q$ is the real photon energy given by Eq.~\eqref{q-W},
and $\gamma_X$ is the real binding momentum. 
The dependence of $F$ on the real energy $W$ is through  $k^2$  and the energy $q$ of the photon.
In Eq.~\eqref{coeffb}, we have used the conservation of energy in Eq.~\eqref{Econservation} 
 to express $b$ as a sum of four terms that are all order $\delta^2$ or smaller
 when $W$ is order $\delta^2/M_{*0}$. 
Note that the sum of the three coefficients in Eqs.~\eqref{coefficientsabc}  does not depend on the energy:%=======================
\begin{equation}
a+b+c = -\gamma_X^2 + i \mu \Gamma_{*0}.   
 \label{a+b+c}
\end{equation}
%========================
The integral over $x$ in Eq.~\eqref{eetoXgammadx} can be evaluated analytically:
%=======================
\begin{equation}
F(W) = - i\frac{\mu \sqrt{\pi \gamma_X}}{4\pi M_0} q
 \left(\frac{b}{2 c}    \log\frac{\sqrt{a} +\sqrt{a+b+c} +  \sqrt{c}}{\sqrt{a} +\sqrt{a+b+c}  - \sqrt{c}}
 +\frac{\sqrt{a}  -\sqrt{a+b+c}}{\sqrt{c}}  \right) .
 \label{Fanalytic}
\end{equation}
%========================
The loop amplitude in Eq.~\eqref{Fanalytic} is particularly simple in the limit $\Gamma_{*0} \to 0$:
%=======================
\begin{eqnarray}
F(W) =   i \frac{\sqrt{\pi \gamma_X}}{4\pi} 
  \left[ \frac{ (\mu/M_0)^2 q^2 + k^2 + \gamma_X^2}{2 (\mu/M_0)q}
   \log\frac{k+ (\mu/M_0)q  + i\gamma_X}{k- (\mu/M_0)q +  i\gamma_X}  - k + i\gamma_X  \right] ,
 \label{Fanalytic0}
\end{eqnarray}
%========================
where $k = i \sqrt{M_{*0}|W|}$ for $W<0$ and $k = \sqrt{M_{*0}|W|}$ for $W>0$. 

The triangle singularity arises from the logarithm in Eq.~\eqref{Fanalytic}.
The denominator of the  argument of the logarithm vanishes at a complex energy
that approaches the real axis  in the limits $\Gamma_{*0} \to 0$  and $\gamma_X \to 0$.
In these limits, the denominator is zero at the energy for which $k = (\mu/M_0)q$.
The energy $W_\triangle$ at which the triangle singularity occurs can be obtained by solving 
the equation $k = (\mu/M_0)q$, where $k = \sqrt{M_{*0} W}$ and  $q$ is the function of $W$ in Eq.~\eqref{q-W}. 
The solution can be expanded in powers of $\delta/(M_{*0}\!+\!M_0)$ and $E_X/\delta$:\\
%=======================
\begin{equation}
W_\triangle \approx \frac{(\mu/M_0)\delta^2}{M_{*0}\!+\!M_0} 
\left( 1 + \frac{\delta^2}{(M_{*0}\!+\!M_0)^2} - \frac{2E_X}{\delta} + \ldots \right).
    \label{Wtriangle}
\end{equation}
%========================
The prediction  from the leading term is $W_\triangle=2.7$~MeV.

%%%%%%%%%%%%%%%%%%%%%%%%%%%%%%%%%%%%%%%%%%%%%%%%
\begin{figure}[t]
\includegraphics*[width=0.7\linewidth]{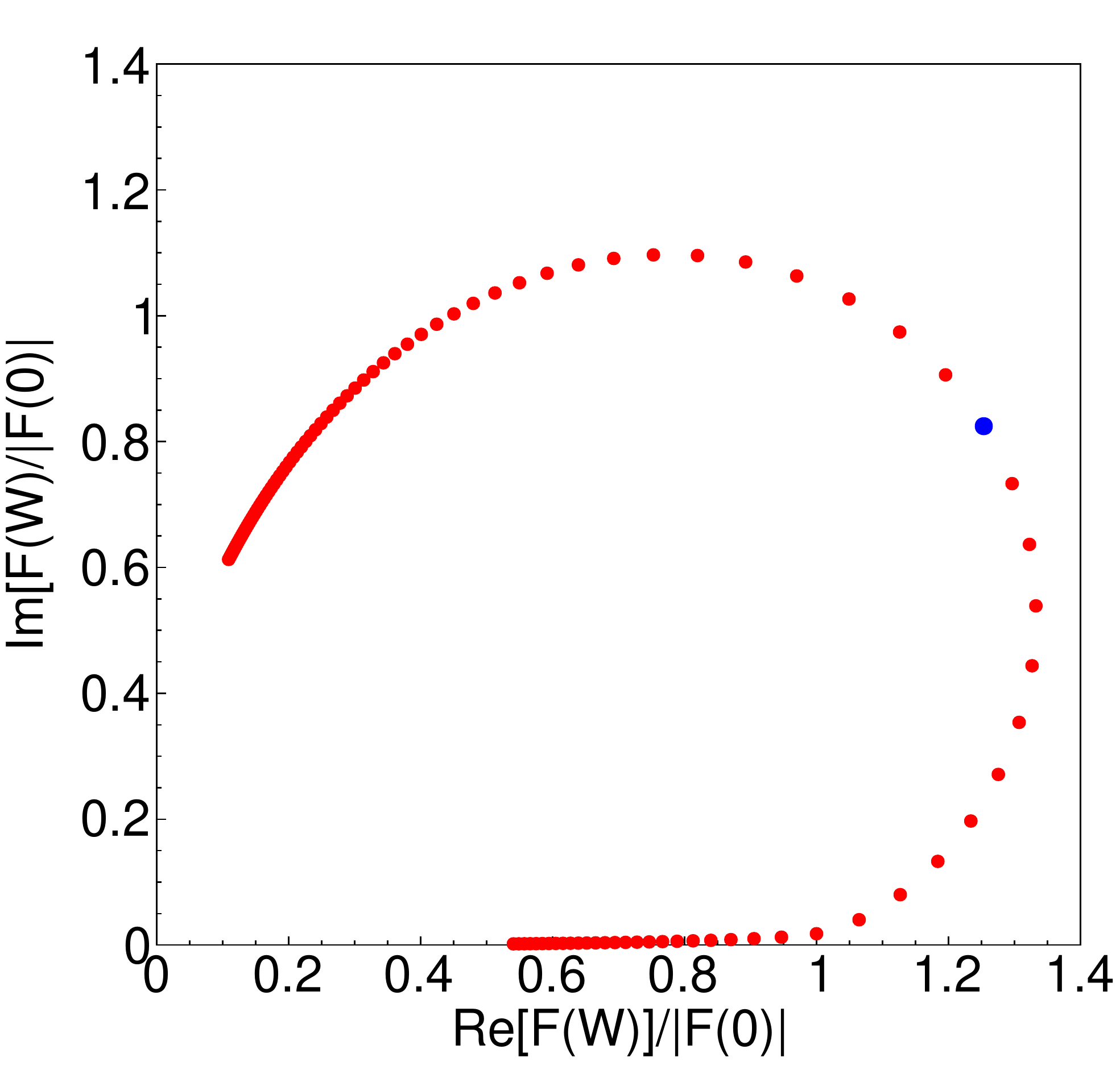} 
\caption{
Argand diagram for the amplitude $F(W)$ for $E_X = -0.17$~MeV.
The dots correspond to  values of $W$ spaced by 0.2~MeV,
and they move counterclockwise as $W$ increases.
The larger (blue) dot is at the value $W=2.2$~MeV that maximizes $|F(W)|^2$.
}
\label{fig:ArgandF}
%\vspace*{0.0cm}
\end{figure}
%%%%%%%%%%%%%%%%%%%%%%%%%%%%%%%%%%%%%%%%%%%%%%

The Argand diagram for the amplitude $F(W)$ at equally spaced values of $W$ is shown in Fig.~\ref{fig:ArgandF}.
As $W$ increases towards 0 from below, $F(W)$ increases along the positive real axis.
As $W$ passes through 0, $\mathrm{Im}[F(W)]$ begins to increase.
The amplitude $F(W)$ then follows a roughly circular path.
The value of $F(W)$ is $(1.20+0.86\,i)\, |F(0)|$ at $W=2.2$~MeV,
where $|F(W)|^2$ has its maximum value.
At large $W$, the decreasing amplitude $F(W)$ approaches the positive imaginary axis.  
Thus the amplitude $F(W)$  moves counterclockwise around a  loop in one quadrant of the complex plane. 
This should be compared to the path followed by the amplitude $A(E)$ for an ideal resonance as a function of the energy $E$. As $E$ approaches the resonance, $A(E)$ increases from 0 along the positive real axis.
It moves counterclockwise around a circle in the upper half of the complex plane,  
crossing the positive imaginary axis as $E$ passes through the maximum of  $|A(E)|^2$.
As $E$ increases further, $A(E)$ decreases towards 0 along the negative real axis.  
The path of the triangle-singularity amplitude is qualitatively similar to that of an ideal resonance,
except that $F(W)$ traces out a loop that remains entirely in one quadrant of the complex plane.

%%%%%%%%%%%%%%%%%%%%%%%%%%%%%%%%%%%%%%%%%%%%%%%%
\begin{figure}[t]
\includegraphics*[width=0.7\linewidth]{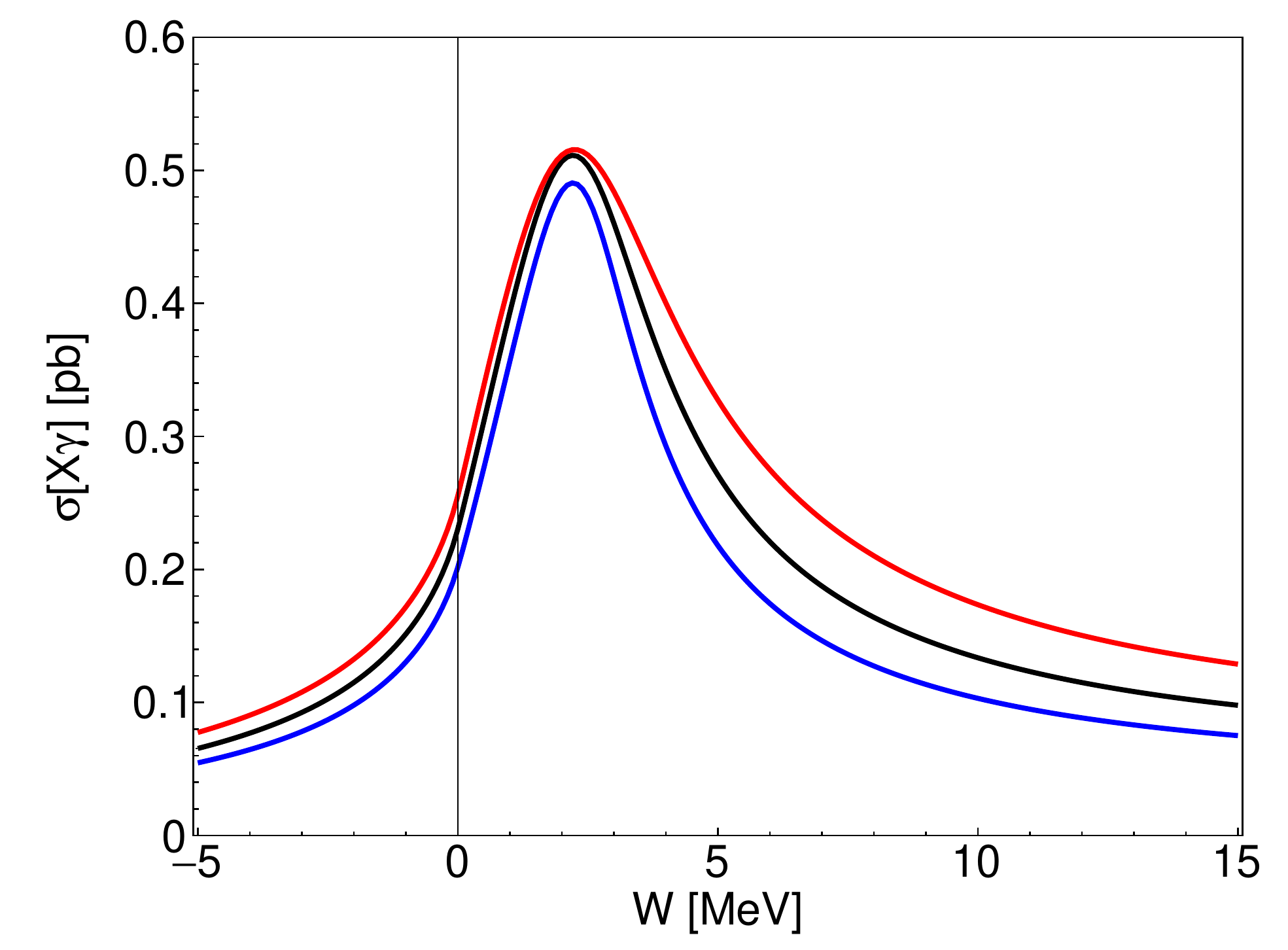} 
\caption{
Cross section for $e^+ e^- \to X(3872)\,  \gamma$ as a function of the 
center-of-mass energy $W$ relative to the $D^{*0} \bar D^{*0}$ threshold.
The cross sections were calculated using the analytic result for  $F(W)$ in Eq.~\eqref{Fanalytic}.
The  three  curves in order of decreasing cross sections are for binding energies 
$|E_X| = 0.30$~MeV (red), 0.17~MeV (black), and 0.10~MeV (blue).
The normalizations of the curves correspond to the amplitudes in Eq.~\eqref{A^2,A2/A0}.
If $A_2/A_0$ is allowed to vary with $|A_0|^2 + |A_2|^2$ fixed,
the normalizations can change by a factor ranging from 0.47 to 1.80.
}
\label{fig:sigma-bindingE}
%\vspace*{0.0cm}
\end{figure}
%%%%%%%%%%%%%%%%%%%%%%%%%%%%%%%%%%%%%%%%%%%%%%

The cross section for $e^+e^- \to X \gamma$ as a function  of the  invariant mass $\sqrt{s}$
is obtained by inserting the loop amplitude $F(W)$ 
in Eq.~\eqref{Fanalytic} into the expression in Eq.~\eqref{sigmagammaXana}.
The cross section near the $D^{*0} \bar D^{*0}$ threshold is shown  in Fig.~\ref{fig:sigma-bindingE} 
for  three values of the binding energy: $|E_X| = 0.30$~MeV, 0.17~MeV, and 0.10~MeV.
The peaks of the line shapes are produced by the triangle singularity. 
The position of the peak is 2.2~MeV above the $D^{*0} \bar D^{*0}$ threshold,
and it is insensitive to the binding energy.
The height of the peak  is also insensitive to $|E_X|$ provided  $|E_X| \gg \Gamma_{*0}$.
In Fig.~\ref{fig:sigma-bindingE}, we have used the preferred
values of the amplitudes $A_0$ and $A_2$ given by Eq.~\eqref{A^2,A2/A0}.
For $|E_X| = 0.17$~MeV, the cross section at the peak is  0.51~pb. 
If we allow all possible complex values of $A_2/A_0$
consistent with the value of  $|A_0|^2 + |A_2|^2$ in Eq.~\eqref{A^2,A2/A0},
the cross section can be larger by a factor of 1.80 or smaller by a factor of 0.47.
Beyond the peak, the cross section decreases to a local minimum at an energy $W$ near 40~MeV 
before increasing because of the $k^3$ dependence of the P-wave cross section for producing $D^{*0}\bar D^{*0}$.
For $|E_X| = 0.17$~MeV, the cross section at the local minimum is about  0.07~pb. 

The BESIII collaboration has measured cross sections for $e^+ e^-$ annihilation into $X\gamma$
at center-of-mass energies ranging from 4.008~GeV to 4.6~GeV 
by observing $X$ in the final states $J/\psi\, \pi^+\pi^-$ and $J/\psi\, \pi^+\pi^-\pi^0$ \cite{Ablikim:2013dyn,Ablikim:2019zio}.
They  did not measure the cross section at energies  between  4.009~MeV and 4.178~MeV,
which includes the predicted energy 4.016~GeV of the peak from the triangle singularity.
The BESIII collaboration measured the cross sections in 10~MeV steps 
between 4.178~GeV and 4.278~GeV \cite{Ablikim:2019zio}.
The largest measured value of the product $\sigma\,\mathrm{Br}$ of the cross section 
and the branching fraction Br of $X$ into  $J/\psi\, \pi^+\pi^-$ was about 0.5~pb.
We derived upper and lower bounds on the branching fraction Br for the $X$ bound state in Ref.~\cite{Braaten:2019ags}.
The BaBar collaboration has recently measured the inclusive branching fraction of $B^+$ into $K^+$
 plus the  $X$ resonance feature \cite{Wormser}.  It implies a branching fraction into  $J/\psi\, \pi^+\pi^-$
from the $X$ resonance feature  that
 provides the loose lower bound $\mathrm{Br} > 4\%$ on the branching fraction from the $X$ bound state.
An upper bound $\mathrm{Br} < 33\%$ can be derived from measurements of branching ratios 
of $J/\psi\, \pi^+\pi^-$ over other short-distance decay modes of the $X$.
Thus the height of the peak from the charm-meson triangle singularity could be 
a significant fraction of the cross section that has been  measured in the higher energy region.

%\newpage

%%%%%%%%%%%%%%%%%%%%%%%%%%%%%%%%
\section{Bound-State Wavefunction}
\label{sec:Wavefunction}
%%%%%%%%%%%%%%%%%%%%%%%%%%%%%%%

In Ref.~\cite{Dubynskiy:2006cj}, Dubynskiy and Voloshin (DV) presented
an approximation for the absorptive contribution to the cross section for $e^+e^- \to  X(3872) \gamma$
that involved the Schr\"odinger wavefunction of the $X$.
The momentum-space wavefunction $\psi(k)$ in the rest frame of the bound state
is a function of the relative momentum $k$ of the constituents.  The standard  normalization for the wavefunction is
%===============
\begin{equation}
\int \frac{d^3k}{(2 \pi)^3}\, |\psi(k)|^2 = 1.
\label{psi-norm}
\end{equation}
%===============
The  universal momentum-space wavefunction for a weakly bound S-wave molecule
whose constituents have short-range interactions is  \cite{Braaten:2004rn} 
%===============
\begin{equation}
\psi_X(k) = \frac{ \sqrt{8 \pi \gamma_X}}{k^2 + \gamma_X^2}.
\label{psiX-k}
\end{equation}
%===============
In the case of the $X$, this wavefunction should be a good approximation out to momenta
$k$ of order $m_\pi$, beyond which it should decrease more rapidly with $k$.

An  expression for the loop amplitude $F(W)$ involving a wavefunction can be obtained  
by closing the contour for the integral over $\omega$ in   Eq.~\eqref{Fd4k}  
 in the lower half-plane. The resulting  loop amplitude  can be expressed as
%===========
\begin{eqnarray}
 F(W)&=& 
- \frac{\mu/M_0}{\sqrt{2}\, M_{*0}}
 \int\!\! \frac{d^3k}{(2\pi)^3} \, 
   \frac{(\bm q \cdot \bm k) \, \psi( \bm q-\bm k, \bm k )}{W  - \bm k^2/M_{*0} + i\Gamma_{*0}},
\label{F-psi}
\end{eqnarray}
%===========
where the last factor in the numerator is
%===========
\begin{eqnarray}
\psi( \bm q-\bm k, \bm k ) =
   \frac{\sqrt{8\pi \gamma_X}}{(\gamma_X^2 - i \mu \Gamma_{*0})   + \big( \bm k - (\mu/M_0) \bm q \big)^2}.
\label{psiX2-q}
\end{eqnarray}
%===========
We have simplified the denominator by using the 
conservation of energy in Eq.~\eqref{Econservation} to eliminate $W$.
The function $\psi( \bm q-\bm k, \bm k )$ in Eq.~\eqref{psiX2-q} can be interpreted as the wavefunction 
for a bound state whose constituents have momenta $\bm q-\bm k$ and $\bm k$.
If we set $\Gamma_{*0}=0$, this is just the universal wavefunction 
in Eq.~\eqref{psiX-k} with the relative momentum $k$ replaced by $|\bm k - (\mu/M_0) \bm q|$,
which is a function of the velocity difference $\bm k/M_{*0} - (\bm q - \bm k)/M_0$ of the constituents.
The wavefunction is expected to depend on the velocity difference in a Galilean-invariant effective field theory.
The wavefunction in Eq.~\eqref{psiX2-q} is the appropriate wavefunction for the bound state in a frame 
where its momentum  is $\bm q$.  
A diagrammatic derivation of this wavefunction is presented in the Appendix.
The wavefunction  in Eq.~\eqref{psiX2-q} satisfies the normalization  condition in Eq.~\eqref{psi-norm} 
 if $\gamma_X$ is real and $\Gamma_{*0}=0$.

The  wavefunction  for the $X$ in its rest frame used by DV in Ref.~\cite{Dubynskiy:2006cj} was
%===============
\begin{equation}
\psi_\mathrm{DV}(k) = 
\frac{\sqrt{\Lambda(\Lambda+\gamma_X)}}{\Lambda-\gamma_X} \sqrt{ 8 \pi \gamma_X}
\left(\frac{1}{k^2 + \gamma_X^2} - \frac{1}{k^2 + \Lambda^2} \right),
\label{psiDV-k}
\end{equation}
%===============
where $\Lambda$ is an adjustable parameter.
DV identified the parameter $\gamma_X$ in Eq.~\eqref{psiDV-k} with the binding momentum of $X$.
This identification can be justified rigorously only in the limit $\Lambda \to \infty$.
The DV wavefunction  in Eq.~\eqref{psiDV-k} is  a regularized form of the universal wavefunction $\psi_X(k)$
in Eq.~\eqref{psiX-k} that reduces to $\psi_X(k)$ in the limit $\Lambda \to \infty$.
The universal wavefunction decreases as $1/k^2$ when $k$ is much larger than $\gamma_X$.
The subtraction in the DV wavefunction in Eq.~\eqref{psiDV-k} makes it decrease as $1/k^4$ 
when $k$ is much larger than the momentum scale $\Lambda$. 
At $k=0$, the DV wavefunction is larger than $\psi_X(0)$ by a factor of $1+3\gamma_X/2\Lambda$. 
In Ref.~\cite{Dubynskiy:2006cj}, DV illustrated their results 
using the values $\Lambda = 200$~MeV and $\Lambda=300$~MeV.
The DV wavefunction was used previously by Voloshin in a study of the decays of  $X$ into 
$D^0 \bar D^0 \gamma$ \cite{Voloshin:2005rt}.

In Ref.~\cite{Dubynskiy:2006cj}, DV assumed that the wavefunction for $X$ in a frame 
where it has momentum $\bm q$ could be obtained from Eq.~\eqref{psiDV-k}
simply by replacing the relative  momentum $k$ by $|\bm k - \tfrac12 \bm q|$,
which is a function of the momentum difference $\bm k - (\bm q -\bm k)$ of the constituents.
DV used that wavefunction to calculate the absorptive contribution to the loop amplitude $F(W)$.
Their prescription for the wavefunction can be used to calculate the full  amplitude
$F_\mathrm{DV}(W)$ by replacing  $\psi(\bm q-\bm k, \bm k)$ in Eq.~\eqref{F-psi}
by $ \psi_\mathrm{DV}(|\bm{k} - \tfrac12 \bm{q}| )$.
The amplitude $F_\mathrm{DV}(W)$ can be obtained most easily by first calculating the amplitude $F_X(W)$
obtained by replacing $\psi(\bm q-\bm k, \bm k)$ in Eq.~\eqref{F-psi}
 by $ \psi_X(|\bm{k} - \tfrac12 \bm{q}| )$, where $\psi_X$ is the universal wavefunction  in Eq.~\eqref{psiX-k}.
The expression for $F_X(W)$ as a Feynman-parameter integral
is the same as Eq.~\eqref{eetoXgammadx} except that $a$, $b$, and $c$ are
%============
\begin{subequations}
\begin{eqnarray}
a &=&   k^2+i M_{*0}\Gamma_{*0} ,
    \\
b &=&  -\big[ \tfrac14   q^2 +   k^2 + \gamma_X^2  \big] - i M_{*0}\Gamma_{*0} ,
\label{coeffb-wf}
    \\
c &=&  \tfrac14  q^2 ,
\end{eqnarray}
\label{coefficientsabc-wf}%
\end{subequations}
%===========
where $k^2=M_{*0}W$ and $q$ is the photon energy given by Eq.~\eqref{q-W}.
These coefficients can be obtained from those in Eqs.~\eqref{coefficientsabc}
by setting $\mu/M_0$ to $\tfrac12$ in the coefficients of $q^2$ in $b$ and $c$ 
and by setting $\mu/M_0$ to 1 in the coefficient of $M_{*0} \Gamma_{*0}$ in $b$.
The analytic result for $F_X(W)$ in the limit $\Gamma_{*0} \to 0$ is
%=======================
\begin{equation}
F_X(W) =   i \frac{\mu \sqrt{\pi \gamma_X}}{2\pi M_0} 
  \Bigg[ \frac{\tfrac14 q^2 + k^2 + \gamma_X^2}{q}
  \log\frac{k +\tfrac12 q  + i\gamma_X}{k -\tfrac12 q + i\gamma_X}  
 - k + i\gamma_X \Bigg] .
 \label{FXanalytic0}
\end{equation}
%========================
The loop amplitude $F_\mathrm{DV}(W)$ in the limit $\Gamma_{*0} \to 0$ can now be obtained by subtracting 
from $F_X(W)$ the expression with $\gamma_X$ replaced by $\Lambda$ except in the factor  $\sqrt{\pi \gamma_X}$, 
and then multiplying by the first prefactor in Eq.~\eqref{psiDV-k}:
%=======================
\begin{eqnarray}
F_\mathrm{DV}(W) &=&   i \frac{\sqrt{\Lambda(\Lambda+\gamma_X)}}{\Lambda-\gamma_X}
\frac{\mu \sqrt{\pi \gamma_X}}{2\pi M_0} 
  \Bigg[ \frac{\tfrac14 q^2 + k^2 + \gamma_X^2}{q}
  \log\frac{k +\tfrac12 q  + i\gamma_X}{k -\tfrac12 q + i\gamma_X}  
+ i\gamma_X  
\nonumber\\
&&\hspace{4.5cm} 
-  \frac{\tfrac14 q^2 + k^2 + \Lambda^2}{q}
  \log\frac{k +\tfrac12 q  + i\Lambda}{k -\tfrac12 q + i \Lambda}  
- i \Lambda\Bigg] .
 \label{FanalyticDV}
\end{eqnarray}
%========================

%%%%%%%%%%%%%%%%%%%%%%%%%%%%%%%%%%%%%%%%%%%%%%%%
\begin{figure}[t]
\includegraphics*[width=0.7\linewidth]{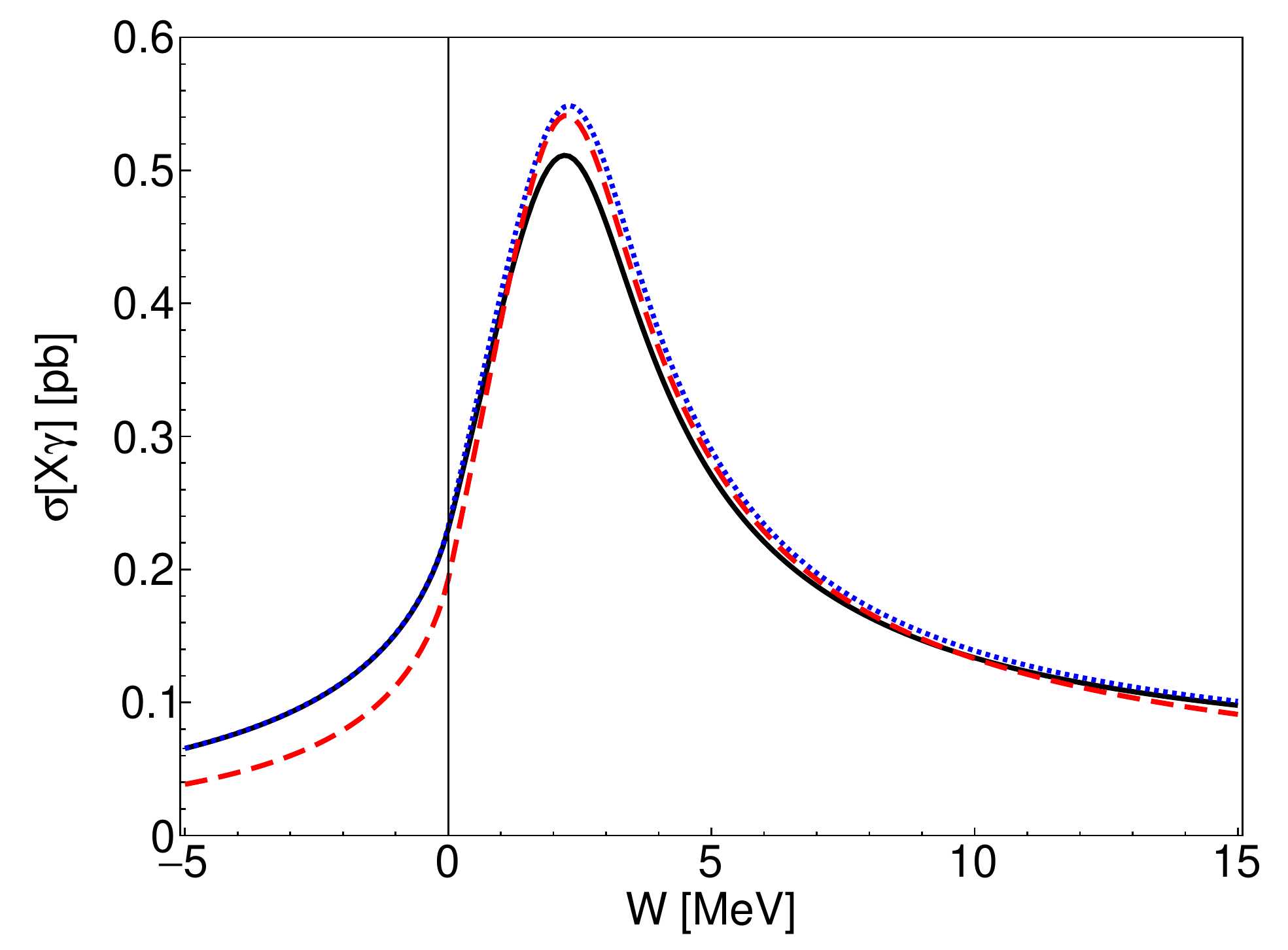} 
\caption{
Cross section for $e^+ e^- \to X  \gamma$ as a function of the 
center-of-mass energy $W$ relative to the $D^{*0} \bar D^{*0}$ threshold.
The binding energy is $|E_X| = 0.17$~MeV.  The cross sections were calculated using 
the complete result for $F(W)$ in Eq.~\eqref{Fanalytic} (solid black curve), 
the approximation for $F(W)$ in Eq.~\eqref{Fanalytic0}  
obtained by setting $\Gamma_{*0}=0$ (dotted blue curve), 
and the approximation $F_\mathrm{DV}(W)$ in Eq.~\eqref{FanalyticDV} with $\Lambda = 200$~MeV (dashed red curve).
}
\label{fig:sigma-X}
%\vspace*{0.0cm}
\end{figure}
%%%%%%%%%%%%%%%%%%%%%%%%%%%%%%%%%%%%%%%%%%%%%%

In Fig.~\ref{fig:sigma-X}, we show the cross sections for $e^+ e^- \to X \gamma$
in the triangle singularity region calculated using the approximation for the loop amplitude 
$F_\mathrm{DV}(W)$ in Eq.~\eqref{FanalyticDV} with $\Lambda = 200$~MeV.
We have set the binding energy to $|E_X| = 0.17$~MeV.
The cross section is compared to the cross sections calculated 
using the complete result for $F(W)$ in Eq.~\eqref{Fanalytic}
and the result for $F(W)$ with $\Gamma_{*0}=0$  in Eq.~\eqref{Fanalytic0}.
The approximation using the DV wavefunction is significantly lower for $E<0$,
it is a little higher at the peak, and it rapidly approaches the exact result as $W$ increases.

%\newpage

%%%%%%%%%%%%%%%%%%%%%%%%%%%%%%%%
\section{Absorptive contribution}
\label{sec:Absorptive}
%%%%%%%%%%%%%%%%%%%%%%%%%%%%%%%

The loop amplitude $F(W)$ in Eq.~\eqref{Fd4k} has an absorptive contribution that corresponds to 
$e^+e^-$ annihilation into on-shell charm mesons $D^{*0} \bar D^{*0}$ followed by 
the rescattering of the charm-meson pair into $X(3872)\gamma$.
In the limit $\Gamma_{*0}\to 0$, the  absorptive contribution is simply the imaginary part of $F(W)$.
The imaginary part can be obtained by cutting rules that replace the propagators of the spin-1 charm mesons
 in Eq.~\eqref{Fd4k} by delta functions. 
The absorptive contribution to the loop amplitude in the limit $\Gamma_{*0}\to 0$ is
%===========
\begin{eqnarray}
\mathrm{Im} \big[F(W)\big]&=&  \frac{\pi \sqrt{\pi \gamma_X} }{M_0}
    \int\!\! \frac{d^3k}{(2\pi)^3} (\bm q \cdot \bm k)\,  \delta\big(\bm k^2 - M_{*0} W\big)
\nonumber\\
&&\hspace{3cm}
\times   \frac{1}{\bm k^2/(2M_{*0}) +(\bm k -\bm q)^2/(2M_0)  +( |\bm q| - \delta) - W - i \epsilon},
%\label{ImFabs-integral}
\end{eqnarray}
%===========
where $W = \sqrt{s}-2M_{*0}$.
The integral over $\bm{k}$ can  be evaluated analytically:
%===========
\begin{eqnarray}
\mathrm{Im} \big[F(W)\big]= \frac{\sqrt{\pi \gamma_X}}{4\pi} k
\left( \frac{ (\mu/M_0)^2 q^2 + k^2 + \gamma_X^2}{4(\mu/M_0) qk}  \log\frac{[(\mu/M_0)q + k]^2 + \gamma_X^2}{[(\mu/M_0)q - k]^2 + \gamma_X^2} - 1 \right) \theta(W),
\label{ImFabs-analytic}
\end{eqnarray}
%===========
where $k = \sqrt{M_{*0}W}$. This expression, which is nonzero only for $W>0$, agrees with the imaginary part of 
the expression for $F(W)$ in Eq.~\eqref{Fanalytic0}, which was obtained by taking the limit $\Gamma_{*0}\to 0$.

%%%%%%%%%%%%%%%%%%%%%%%%%%%%%%%%%%%%%%%%%%%%%%%%
\begin{figure}[t]
\includegraphics*[width=0.7\linewidth]{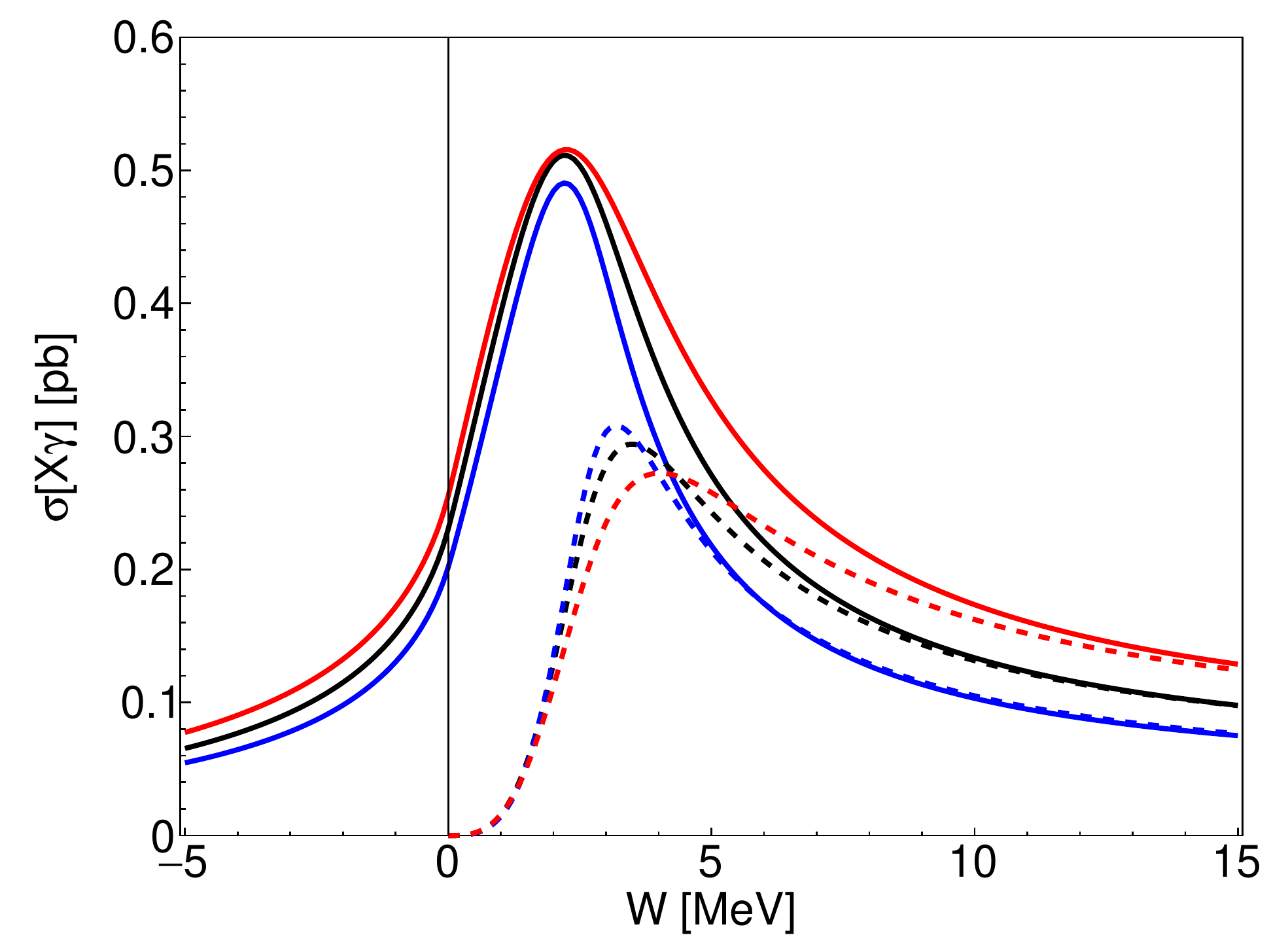} 
\caption{
Cross section for $e^+ e^- \to X \gamma$ as a function of the 
center-of-mass energy $W$ relative to the $D^{*0} \bar D^{*0}$ threshold.
The solid curves in order of decreasing cross sections are for binding energies 
$|E_X| = 0.30$~MeV, 0.17~MeV, and 0.10~MeV.
The dashed curves are the absorptive contributions, 
which approach the corresponding cross sections as $W$ increases.
}
\label{fig:sigma-absorp}
%\vspace*{0.0cm}
\end{figure}
%%%%%%%%%%%%%%%%%%%%%%%%%%%%%%%%%%%%%%%%%%%%%%

In Fig.~\ref{fig:sigma-absorp}, we compare the  cross section for $e^+ e^- \to X  \gamma$  
in the triangle singularity region  with the absorptive contribution obtained by replacing $F(W)$ 
in Eq.~\eqref{sigmagammaXana} by $\mathrm{Im} [F(W)]$ in Eq.~\eqref{ImFabs-analytic}.
The absorptive contribution is zero below the $D^{*0} \bar D^{*0}$ threshold.
Unlike the complete cross section, the position of the peak of the absorptive contribution depends 
on the binding energy. 
For $|E_X|=0.17$~MeV, the position of the peak is about $1.3$~MeV higher than that of the full cross section, and
the height of the peak is about 58\% of that of the  full cross section. 
Thus the absorptive contribution is not a good approximation to the cross section
for $e^+e^- \to X \gamma$ in the triangle singularity region.
At larger energies, the absorptive contribution quickly approaches the full cross section.

In Ref.~\cite{Dubynskiy:2006cj}, Dubynskiy and Voloshin 
derived an approximation for the absorptive contribution to the cross section for $e^+e^- \to  X \gamma$
using a Schr\"odinger wavefunction, as discussed in Section~\ref{sec:Wavefunction}.
Their approximation for the imaginary part of the loop amplitude $F(W)$ 
in Eq.~\eqref{Fd4k} can be expressed as 
%===========
\begin{eqnarray}
\mathrm{Im}\big[ F_\mathrm{DV}(W)\big]=
   \frac{ \pi  \mu}{ \sqrt{2}M_0}  \int\!\! \frac{d^3k}{(2\pi)^3} \, 
 (\bm q \cdot \bm k)\,  \delta\big(\bm k^2 - M_{*0} W\big) \,  
 \psi_\mathrm{DV}\big(\big|\bm{k} - \tfrac12 \bm{q}\big| \big)  ,
\label{ImFDV-int}
\end{eqnarray}
%===========
where  $\psi_\mathrm{DV}$ is the  DV wavefunction in Eq.~\eqref{psiDV-k}.
The integral over $\bm{k}$ in Eq.~\eqref{ImFDV-int}  can  be evaluated analytically.
If we replace $\psi_\mathrm{DV}$ in Eq.~\eqref{ImFDV-int} 
by the universal wavefunction $\psi_X$  in Eq.~\eqref{psiX-k}, the imaginary part of $F(W)$ becomes
%===========
\begin{eqnarray}
\mathrm{Im} \big[F_X(W)\big]= \frac{\mu \sqrt{\pi \gamma_X} }{2 \pi M_0} k
\left[ \frac{\tfrac14 q^2+k^2+\gamma_X^2}{2qk}  
\log\frac{\big(\tfrac12 q+k\big)^2+\gamma_X^2}{\big(\tfrac12 q-k\big)^2+\gamma_X^2} - 1 \right]
\theta(W).
\label{ImFX-analytic}
\end{eqnarray}
%===========
This agrees with the absorptive part in Eq.~\eqref{ImFabs-analytic} if $\mu/M_0=0.518$ is set to $\tfrac12$.
The corresponding integral with the DV wavefunction in Eq.~\eqref{psiDV-k}
can be obtained by subtracting from the factor in square brackets the corresponding factor with
$\gamma_X$ replaced by $\Lambda$ and then multiplying by the first prefactor in Eq.~\eqref{psiDV-k}:
%===========
\begin{eqnarray}
\mathrm{Im} \big[F_\mathrm{DV}(W)\big] &=& 
\frac{\sqrt{\Lambda(\Lambda+\gamma_X)}}{\Lambda-\gamma_X}
\frac{\mu \sqrt{\pi \gamma_X}}{2 \pi M_0} k
\Bigg[ \frac{\tfrac14 q^2+k^2+\gamma_X^2}{2qk}  
\log\frac{\big(\tfrac12 q+k\big)^2+\gamma_X^2}{\big(\tfrac12 q-k\big)^2+\gamma_X^2} 
\nonumber\\
&&\hspace{4cm}
- \, \frac{\tfrac14 q^2+k^2+\Lambda^2}{2qk}  
\log\frac{\big(\tfrac12 q+k\big)^2+\Lambda^2}{\big(\tfrac12 q-k\big)^2+\Lambda^2}  \Bigg] \, \theta(W).
\label{ImFDV-analytic}
\end{eqnarray}
%===========

In Ref.~\cite{Dubynskiy:2006cj},  Dubynskiy and Voloshin estimated the peak in the absorptive cross section 
to be ``numerically of the order of 1~pb''.
They used as input a measurement of the cross section for $e^+ e^- \to D^{*0} \bar D^{*0}$ by the CLEO-c collaboration.
They quoted the cross section as 0.15~nb at a center-of-mass energy 1.6~MeV above the $D^{*0} \bar D^{*0}$ threshold.
This measurement does not seem to appear in the conference proceeding they gave as a reference \cite{Poling:2006da}.
If we insert that cross section into Eq.~\eqref{A0,A2},
we get $|A_0|^2+|A_2|^2= 410~\mathrm{GeV}^{-2}$.
This  differs only by  a factor of about 1.5 from 
the value in Eq.~\eqref{A^2,A2/A0} that we obtained from a fit to the cross section for $e^+ e^- \to D^{*+} \bar D^{*-}$.

%\newpage

%%%%%%%%%%%%%%%%%%%%%%%%%%%%%%%%%%%%
\section{Summary}
\label{sec:Summary}
%%%%%%%%%%%%%%%%%%%%%%%%%%%%%%%%%%%%

In this paper, we presented details of the calculation of the cross section for $X(3872)\gamma$ from  $e^+ e^-$ annihilation. 
A pair of P-wave spin-1 neutral charm mesons is created by the virtual photon from $e^+ e^-$ annihilation, 
and the charm mesons then rescatter into $X\gamma$.
The cross section for producing $X\gamma$ is given in Eq.~\eqref{sigmagammaXana},
and the loop amplitude $F(W)$ is given in Eq.~\eqref{Fanalytic}.
We predicted the normalization of the cross section by using a previous fit to the cross sections for 
$e^+ e^- \to D^{*} \bar D^{*}$ from Belle data  by Uglov {\it et al.}\  \cite{Uglov:2016orr}
to determine the amplitudes $A_0$ and $A_2$ in Eq.~\eqref{A^2,A2/A0}.
The cross section has a narrow peak at an energy  2.2~MeV above the $D^{*0} \bar D^{*0}$ threshold 
 as shown in Fig.~\ref{fig:sigma-bindingE}.  
 We presented the cross section in this region in a previous paper  \cite{Braaten:2019gfj}. 
The peak is caused by a charm-meson triangle singularity.
The height of the peak is predicted to be between 0.2~pb and 0.9~pb if the amplitude ratio $A_2/A_0$
is allowed to vary with $|A_0|^2 + |A_2|^2$ fixed at the value in Eq.~\eqref{A^2,A2/A0}.
If the $X$ is observed in the decay mode $J/\psi\, \pi^+\pi^-$, the cross section must be multiplied by the 
branching fraction Br for the bound state to decay into $J/\psi\, \pi^+\pi^-$.
The loose lower bound $\mathrm{Br} > 4\%$ and the upper bound $\mathrm{Br} < 33\%$
on that branching fraction were derived in Ref.~\cite{Braaten:2019ags}.

The Argand diagram for the loop amplitude from the triangle singularity is shown in Fig.~\ref{fig:ArgandF}.
It is qualitatively similar to that of an ideal resonance, with the amplitude tracing out a counterclockwise loop 
in the complex plane as the energy $W$ increases. Unlike the case of an ideal resonance, 
the loop is restricted  to a singe quadrant of the complex plane.
 
The loop amplitude is expressed in terms of the Schr\"odinger wavefunction of the bound state
with nonzero momentum in Eq.~\eqref{F-psi}.
Such an expression was previously used by Dubynskiy and Voloshin 
in their calculation of the production of $X\gamma$ \cite{Dubynskiy:2006cj}. 
Their wavefunction is a function of the relative momentum of the two constituents.  
In the Appendix, we derived the wavefunction for a bound state with nonzero momentum 
from the  transition amplitude for the case of a near-threshold S-wave resonance.
The wavefunction is a function of the relative velocity of the two constituents.

In the previous study of the production of $X\gamma$ from  $e^+ e^-$ annihilation by Dubynskiy and Voloshin,
they calculated only the absorptive contribution to the cross section.
The comparison between the full cross section and the absorptive contribution is shown in Fig. \ref{fig:sigma-absorp}. 
The absorptive contribution is not a good approximation near the triangle singularity region.  
In the absorptive contribution, the widths of the spin-1 charm mesons are set to 0.  
The narrow peak in the cross section comes from the 
triangle singularitiy that arises when all three charm mesons that form the triangle are on shell.
The binding energy of the $X$ prevents both of its constituents from being on shell simultaneously,
and this was taken into account in DV.  However the widths of the spin-1 charm mesons prevent them 
from being on shell, so the widths are also important.

We calculated the cross section for $e^+ e^- \to X\gamma$ using the amplitudes from the 
charm-meson triangle diagrams in Fig.~\ref{fig:eetogammaX}, which begin with the creation of 
a $D^{*0} \bar D^{*0}$ pair.  Dubynskiy and Voloshin have pointed out that  there are also 
short-distance contributions to that amplitude that begin with the creation of 
$D \bar D$, $D^* \bar D$, or $D \bar D^*$ \cite{Dubynskiy:2006cj}.
Those amplitudes will be essentially constant in the region of the peak from the triangle singularity.
If the short-distance amplitudes are larger than the amplitude from the triangle diagrams,
the cross section near the $D^{*+} D^{*-}$ threshold will have only a small peak or even a dip
on top  of a larger smooth background cross section.
We have assumed the short-distance amplitudes are negligible compared to the amplitude from the triangle diagrams.
Quantitative estimates of the short-distance amplitudes would be useful.

In our prediction for the cross section near the peak from the triangle singularity in Fig.~\ref{fig:sigma-absorp},
we assumed the $X$ is a narrow bound state. 
A prescription for calculating the cross section in the case of  a resonance with a different character
is presented in the Appendix.
As illustrated in Fig.~\ref{fig:sigmatri},  the peak from the triangle singularity in the case of a zero-energy resonance 
or a virtual state can be qualitatively similar to that in the case of a narrow bound state.

The BESIII collaboration has measured the cross section for $X\gamma$ from $e^+ e^-$ annihilation
at center-of-mass energies ranging from 4.008~GeV to 4.6~GeV \cite{Ablikim:2013dyn,Ablikim:2019zio}.
The cross section seems to have a broad peak near 4.2 GeV.
The cross section  has not been measured in the region near the  $D^{*0} \bar D^{*0}$ threshold at 4.014~MeV.
The height of the narrow peak near 4.016~MeV from the charm-meson triangle singularity   
is predicted to be large enough that it could be observed by the BESIII detector.
The observation of this peak would provide strong support for the identification of the $X(3872)$ 
as a weakly bound charm-meson molecule.

%%%%%%%%%%%%%%%%%%%%%%%%%%%%%%%%%%%%%%%%%%
\begin{acknowledgments}
% put your acknowledgments here.
This work was supported in part by the Department of Energy under grant DE-SC0011726
and by the National Science Foundation under grant  PHY-1607190.
\end{acknowledgments}
%%%%%%%%%%%%%%%%%%%%%%%%%%%%%%%%%%%%%%%%%%

%\newpage

\appendix

%%%%%%%%%%%%%%%%%%%%%%%%%%%%%%%%%%%%%%%%%%
\section{Aspects of a Near-threshold S-wave Resonance } 
%%%%%%%%%%%%%%%%%%%%%%%%%%%%%%%%%%%%%%%%%%

In this Appendix, we describe some aspects of the physics of the $X(3872)$ that can be deduced from the transition amplitude of a near-threshold S-wave resonance.

%%%%%%%%%%%%%%%%%%%%%%%%%%%%%%%%%%%%%%%%%%
\subsection{Transition amplitude}

\begin{figure}[t]
\includegraphics*[width=0.45\linewidth]{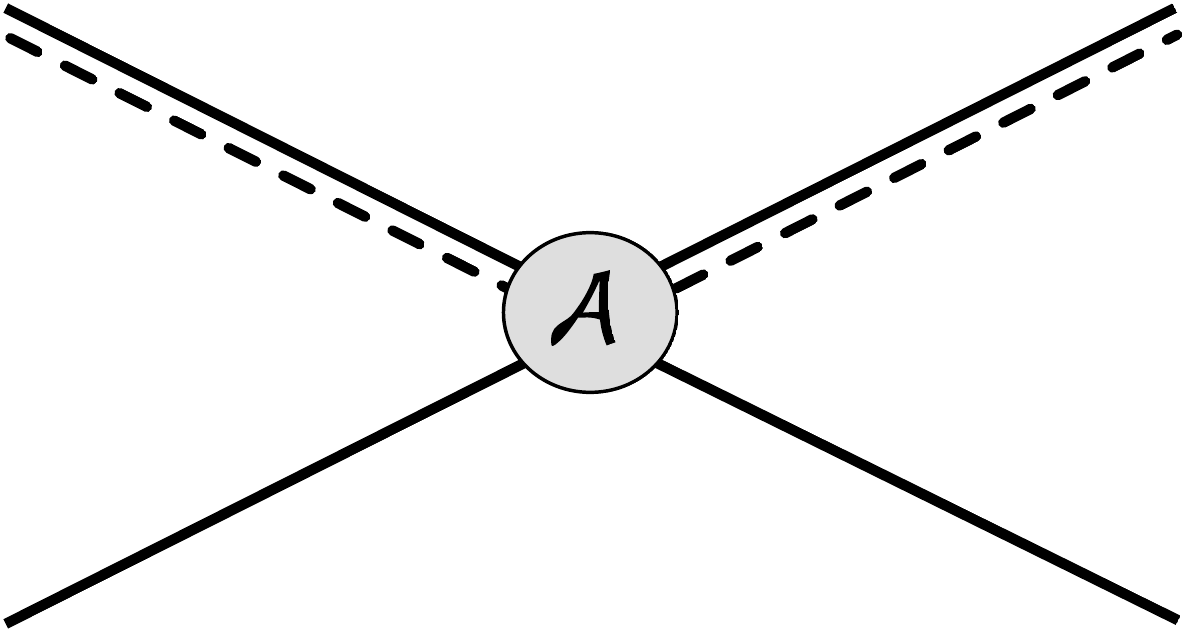} 
\caption{The 1PI transition amplitude $\mathcal{A}$ for the
two particles with a near-threshold S-wave resonance.
The connected $2 \to 2$ transition amplitude is the product of 
the 1PI  amplitude $\mathcal{A}$ and a propagator for each of the 4 external legs.
 }
\label{fig:2to2amplitude}
\end{figure}

We consider a stable particle with mass $M_0$
and a particle with mass $M_{*0}$ and decay width $\Gamma_{*0}$.
The particles have  short-range interactions that produce an S-wave resonance very close to the scattering threshold, 
whose energy  we take to be $E=0$.
Many aspects of the two-body physics can be derived from
the one-particle-irreducible (1PI) transition amplitude for the two particles,
which is illustrated in Fig.~\ref{fig:2to2amplitude}.  
The simplest plausible model for the 1PI transition amplitude as a function of the total energy $E$ of the pair of particles 
in the center-of-momentum (CM) frame  is
%===========
\begin{equation}
\mathcal{A}(E) = 
   \frac{2\pi /\mu}{- \gamma_X  + \sqrt{-2\mu (E + i \Gamma_{*0}/2)}}.
\label{Amp-E}
\end{equation}
%===========
This is obtained from the universal 1PI transition amplitude for a near-threshold S-wave resonance  \cite{Braaten:2004rn}
by (1) shifting the energy threshold for the pair of particles from 0 to $- i \Gamma_{*0}/2$
to take into account the decay width of the second particle 
and (2) generalizing the real inverse scattering length $\gamma_X$ to a complex parameter 
with a positive imaginary part to take into account short-distance decay channels.
The T-matrix element for elastic scattering of the two particles with total CM energy $E$ is obtained by evaluating 
the amplitude in Eq.~\eqref{Amp-E} at that real energy.

%%%%%%%%%%%%%%%%%%%%%%%%%%%%%%%%%%%%%%%%%%%%%%%%
%%%%%%%%%%%%%%%%%%%%%%%%%%%%%%%%%%%%%%%%%%%%%%

Since the pair of particles has a resonance near the scattering threshold,
the scattering amplitude has a pole at an energy $E_\mathrm{pole}$ near the scattering threshold.
The pole energy may be complex if there are inelastic scattering channels.
If $E_\mathrm{pole}$ is on the physical sheet of the complex energy $E$, the resonance is referred to as a {\it bound state}.
It is a  {\it narrow  bound state} if  $\mathrm{Re}[E_\mathrm{pole}]$  
is much larger in absolute value than $\mathrm{Im}[E_\mathrm{pole}]$.
If $E_\mathrm{pole}$ is on another sheet, the resonance is referred to as  a {\it virtual state}.
The transition amplitude in Eq.~\eqref{Amp-E} is an  analytic  function of  the complex energy $E$
with a pole at the energy $E_\mathrm{pole} \equiv E_X' - i \Gamma_X/2$, whose
real and imaginary parts are
%===========
\begin{subequations}
\begin{eqnarray}
E_X'  &= &- \frac{\mathrm{Re}[\gamma_X]^2 - \mathrm{Im}[\gamma_X]^2}{2 \mu}
\label{EX}
\\
\Gamma_X &=& \Gamma_{*0} + \frac{2\, \mathrm{Re}[\gamma_X]\,\mathrm{Im}[\gamma_X]}{\mu}.
\label{GammaX}
\end{eqnarray}
\label{EX,GammaX}%
\end{subequations}
%===========
We have denoted the real part of $E_\mathrm{pole}$ by $E_X'$ to distinguish it from the 
measured energy $E_X$ of the $X$ in Eq.~\eqref{EX-exp},
which can be identified with the center of the resonance in the $J/\psi\, \pi^+\pi^-$ channel.
If $\mathrm{Re}[\gamma_X]>0$, the resonance is a {\it bound state}.
It is a {\it narrow bound state} if $|\mathrm{Re}[\gamma_X]|$ is  much larger than both $\mathrm{Im}[\gamma_X]$
and $\sqrt{\mu \Gamma_{*0}}$.
If $\mathrm{Re}[\gamma_X]<0$, the resonance is a {\it virtual state}.

The behavior of the transition amplitude at complex energies $E$ near the pole is
%===========
\begin{equation}
\mathcal{A}(E) \longrightarrow 
   \frac{-2\pi \gamma_X/\mu^2}{E-(E_X - i \Gamma_X/2)}.
\label{Amp-E:pole}
\end{equation}
%===========
This pole approximation for $\mathcal{A}(E)$ 
is a good approximation for the T-matrix element over some real range of the energy $E$ 
only if the resonance is a narrow bound state.
In this case,  the residue of the pole in Eq.~\eqref{Amp-E:pole} determines the vertex 
for the coupling of the bound state  to a pair of particles: $i \sqrt{2\pi \gamma_X}/\mu$. 
This vertex can be used to calculate production rates of the bound state diagrammatically.
In the case of the $X$, the resonance is in the S-wave channel for the superposition of charm mesons in
Eq.~\eqref{Xflavor}.  The vertices for the coupling of the $X$ to $D^{*0} \bar D^0$  or to $D^0 \bar D^{*0}$
are therefore given by Eq.~\eqref{vertexX}.

%%%%%%%%%%%%%%%%%%%%%%%%%%%%%%%%%%%%%%%%%%
\subsection{Short-distance production}
\label{sec:SDproduction}

If there is a reaction that can produce the two particles at short distances,
the inclusive production rate of the two particles and their decay products can be determined by the optical theorem.  
The amplitude for the production of the two particles can be expressed as the product of $\mathcal{A}(E)$,
which depends on the CM energy of the two particles, and a short-distance factor that is insensitive to $E$.
We first ignore inelastic effects, and assume the T-matrix is exactly unitary.
In the integral of the square of the amplitude over the momenta $\bm{p}_1$ and $\bm{p}_2$  of the two particles,
it is convenient to change variables to the total momentum $\bm{P} = \bm{p_1} + \bm{p}_2$ and $E$:
%===========
\begin{subequations}
\begin{eqnarray}
\big| \mathcal{A}(E) \big|^2\,  \frac{d^3p_1}{(2 \pi)^3} \, \frac{d^3p_2}{(2 \pi)^3} &= &
\big| \mathcal{A}(E) \big|^2\, \frac{\sqrt{2 \mu E}\, \mu\,  dE}{2 \pi^2} \frac{d^3P}{(2 \pi)^3}
\label{rate-phasespace}
\\
 &=& \frac{1}{\pi}\,  \mathrm{Im} \big[\mathcal{A}(E) \big]\, dE\, \frac{d^3P}{(2 \pi)^3}.
\label{rate-optical}
\end{eqnarray}
\label{rate}%
\end{subequations}
%===========
In the last step, we used the optical theorem for a 2-particle system that is exactly unitary.
If the  range of $E$ is extended to negative values, the last expression in Eq.~\eqref{rate}
can take into account the production of bound states.
We now consider inelastic effects.  
If they are taken into account in a way that ensures the positivity of $\mathrm{Im}[\mathcal{A}(E)]$,
the last expression in Eq.~\eqref{rate} can be interpreted as a factor in the inclusive production rate of 
the {\it resonance feature}, which includes the entire contribution enhanced by the resonance.  
In addition to bound states, it includes final states from decays of the two particles as well as from their inelastic scattering.
If there is a near-threshold S-wave resonance,
the resonance feature includes a  peak above the scattering threshold called a {\it threshold enhancement}
from the production of the pair of particles.
The resonance feature may also include additional structure,
such as a narrow peak below the threshold from a bound state or a peak near the threshold from a virtual state.

The imaginary part of the simple  model amplitude in Eq.~\eqref{Amp-E} at a real energy $E$ can be expressed as
%===========
\begin{equation}
\mathrm{Im} \big[\mathcal{A}(E) \big]= 
\frac{\mu}{2 \pi}\big|\mathcal{A}(E) \big|^2
\left(  \mathrm{Im}[\gamma_X]  + \Big[\mu \sqrt{E^2+  \Gamma_{*0}^2/4} + \mu E \Big]^{1/2} \right).
\label{ImAmp-E}
\end{equation}
%===========
The unitarity condition $\mathrm{Im} [\mathcal{A}(E)] \ge 0$ requires the imaginary part of  $\gamma_X$ to be positive.
The first term in Eq.~\eqref{ImAmp-E} proportional to $ \mathrm{Im}[\gamma_X] $ is the contribution 
from decays of the resonance into {\it short-distance-decay channels}, 
whose ultimate final states include particles with large momentum.
In the case of $X$, they include $J/\psi\, \pi^+\pi^-$, $J/\psi\, \pi^+\pi^-\pi^0$, $J/\psi\, \gamma$,
$\psi(2S)\, \gamma$, and $\chi_{c1}(1P)\, \pi^0$.
The second term is the contribution from the {\it constituent-decay channels}. 
In the case of  $X$, their ultimate final states are $D^0 \bar D^0 \pi^0$ and $D^0 \bar D^0 \gamma$.  
This contribution includes a threshold enhancement from production of  a pair of  constituents
followed by the subsequent decay of $D^{*0}$ or $\bar D^{*0}$.
The measured energy $E_X$ of the $X$ in Eq.~\eqref{EX-exp}
can be identified with the center of energy of the $\mathrm{Im}[\gamma_X]$ term in Eq.~\eqref{ImAmp-E}.

In the case of a narrow bound state,  the short-distance production rate has a narrow peak below the threshold.
The position of the peak, the center of energy $E_X$ of the resonance, 
and the real part $E_X'$ of the pole energy in Eq.~\eqref{EX} are all approximately $- \mathrm{Re}[\gamma_X]^2/2\mu$.
By inserting the pole approximation in Eq.~\eqref{Amp-E:pole} for the amplitude  $\mathcal{A}(E)$
into Eq.~\eqref{ImAmp-E}, the inclusive line shape near the peak can be approximated by
%===========
\begin{equation}
\mathrm{Im} \big[\mathcal{A}(E) \big] \approx 
\frac{2 \pi |\gamma_X|^2/\mu^3}{(E-E_X')^2 + \Gamma_X^2/4}
\left(  \mathrm{Im}[\gamma_X]  + \Big[\mu \sqrt{{E_X'}^2+  \Gamma_{*0}^2/4} + \mu E_X' \Big]^{1/2} \right),
\label{ImAmp-Eapprox}
\end{equation}
%===========
where $E_X'$ is given in Eq.~\eqref{EX}.
Inside an integral over the energy, this line shape can be further approximated by a delta function.
Its coefficient can be expanded in powers of $\mathrm{Im}[\gamma_X] / \mathrm{Re}[\gamma_X]$
and $\sqrt{\mu\Gamma_{*0}} / \mathrm{Re}[\gamma_X]$.
Up to relative corrections suppressed by three powers of $1/ \mathrm{Re}[\gamma_X]$,
the line shape can be expressed as 
%===========
\begin{equation}
\mathrm{Im} \big[\mathcal{A}(E) \big] \approx 
\frac{2 \pi^2}{\mu^2} 
\sqrt{\mathrm{Re}[\gamma_X]^2 + 2\,  \mathrm{Im}[\gamma_X]^2}\, \delta(E-E_X') .
\label{ImAmp-Edelta}
\end{equation}
%===========
The square root reduces to $|\gamma_X|$ to leading order in $\mathrm{Im}[\gamma_X] /\mathrm{Re}[\gamma_X]$.
If the resonance is not a narrow bound state, the real part $E_X'$ of the pole energy in Eq.~\eqref{EX} 
is not  a good approximation for the center of energy $E_X$ of the resonance. 
In the virtual-state limit where $\mathrm{Re}[\gamma_X]$ is negative and  much larger 
in absolute value than $\mathrm{Im}[\gamma_X]$ and $\sqrt{\mu \Gamma_{*0}}$,
both $E_X'$ and $E_X$ are order $ \mathrm{Re}[\gamma_X]^2/\mu$,
but $E_X'$ is negative while $E_X$ is positive.

%%%%%%%%%%%%%%%%%%%%%%%%%%%%%%%%%%%%%%%%%%%%%%%%
\begin{figure}[t]
\includegraphics*[width=0.8\linewidth]{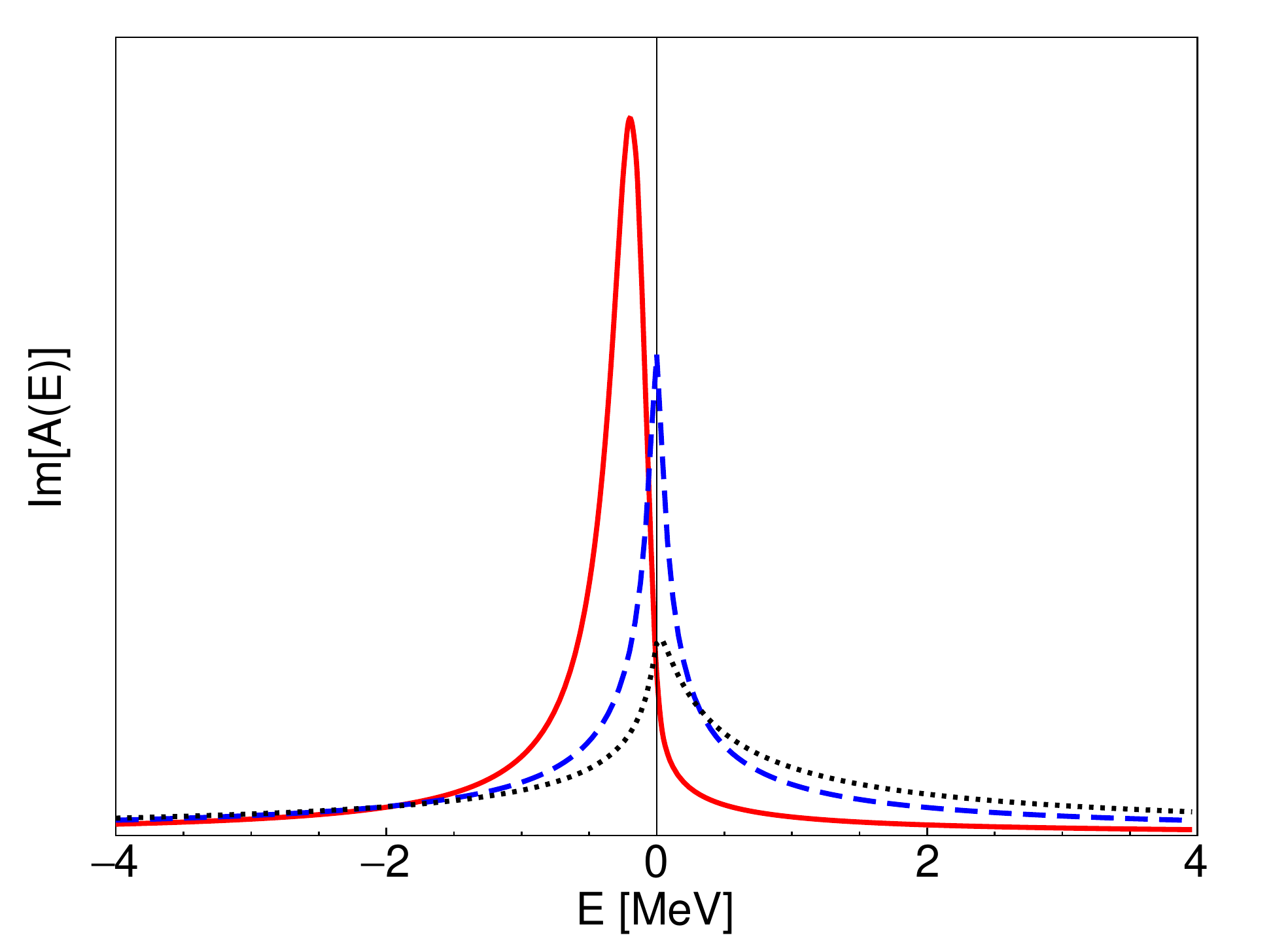} 
\caption{Short-distance-decay contribution to the line shape $\mathrm{Im}[\mathcal{A}(E)]$ 
of a near-threshold S-wave resonance as a function of the energy $E$ in the CM frame.
The curves are for a narrow bound state with $E_X = -0.30$~MeV (solid red), 
a zero-energy resonance with $E_X =0$ (dashed blue), 
and a virtual state with $E_X = +0.30$~MeV (dotted black).
The normalizations of the curves are arbitrary, but the areas under the three curves are equal.
}
\label{fig:lineshape}
\end{figure}
%%%%%%%%%%%%%%%%%%%%%%%%%%%%%%%%%%%%%%%%%%%%%%

In Ref.~\cite{Braaten:2019ags}, a simple model for the line shapes was used to 
illustrate different possibilities for the character of the resonance.
The model for the transition amplitude is $\mathcal{A}(E)$ in Eq.~\eqref{Amp-E},
which depends on $\Gamma_{*0}  = 56$~keV and the two adjustable parameters
$\mathrm{Re}[\gamma_X]$ and $\mathrm{Im}[\gamma_X]$.
The inclusive production rate for the resonance feature  is proportional to
$\mathrm{Im}[\mathcal{A}(E)]$ in Eq.~\eqref{ImAmp-E}.
The contribution from a specific short-distance decay mode $i$ is $\gamma_i \, |\mathcal{A}(E)|^2$,
where $\gamma_i$ is an adjustable normalization factor proportional to the branching fraction
 into that decay mode.  The total production rate of the resonance in that decay mode is
 obtained by integrating over the energy $E$.  The integral over the energy range 
 $E_\mathrm{min} < E < E_\mathrm{max}$ depends logarithmically on the endpoints.
 To define a model with a finite production rate, we follow Ref.~\cite{Braaten:2019ags}
 in declaring the resonance to be the energy region between specified endpoints.
 Our model for the line shape in the short-distance decay mode  $i$ is therefore
 %===========
\begin{equation}
\mathrm{Im} \big[\mathcal{A}(E) \big]\Big|_i= \gamma_i\, \big|\mathcal{A}(E) \big|^2\, 
\Theta\big( E_\mathrm{min} < E < E_\mathrm{max} \big).
\label{lsh-model}
\end{equation}
 %===========
We also follow Ref.~\cite{Braaten:2019ags} in choosing the endpoints to be the 
$D^0 \bar D^0 \pi^0$ threshold $E_\mathrm{min}  = -7.0$~MeV 
and the $D^{*+} D^-$ threshold $E_\mathrm{max} = +8.2$~MeV.
We identify the measured energy $E_X$ of the  $X$ resonance in Eq.~\eqref{EX-exp}
with the center of resonance in the decay mode $i=J/\psi\, \pi^+\pi^-$.
The center of energy $E_X$ for the line shape in Eq.~\eqref{lsh-model}
can be defined by the condition
%===========
\begin{equation}
\int_ {E_\mathrm{min}}^{E_X} \big|\mathcal{A}(E) \big|^2
\,=\, \int_ {E_X}^{E_\mathrm{max}} \big|\mathcal{A}(E) \big|^2.
\label{centerofE}
\end{equation}
 %===========
If $\mathrm{Im}[\gamma_X]$ is fixed, $\mathrm{Re}[\gamma_X]$ can be adjusted 
to get a specified value of $E_X$.
The production rate in the decay channel $i$ is proportional to 
$\gamma_i\,\int_ {E_\mathrm{min}}^{E_\mathrm{max}} |\mathcal{A}(E)|^2$.
This can be made independent of $E_X$ by adjusting the prefactor $\gamma_i$  as a function of $E_X$. 
We choose a fixed value for the imaginary part of  $\gamma_X$:
$\mathrm{Im}[\gamma_X]= \sqrt{\mu \Gamma_{*0}} = 7.4$~MeV.
We consider three cases with different values of the center of energy $E_X$:
\begin{itemize}
\item
{\it narrow bound state}: $E_X = -0.30$~MeV, which requires $\mathrm{Re}[\gamma_X] = 18.8$~MeV. 
This is a little smaller than the  value of  $\gamma_X$ for a narrow bound state 
predicted from Eq.~\eqref{EX-gammaX}:  $\sqrt{2\mu |E_X|}= 24.1$~MeV.  
We use this case to define a dimensionless normalization factor $\hat \gamma_i=1$.
\item
{\it zero-energy resonance}: $E_X =0$, which requires $\mathrm{Re}[\gamma_X] = -6.1$~MeV.
The dimensionless normalization factor is $\hat \gamma_i=2.48$. 
\item
{\it virtual state}: $E_X = +0.30$~MeV, which requires $\mathrm{Re}[\gamma_X] = -27.2$~MeV. 
The dimensionless normalization factor is $\hat \gamma_i=4.20$. 
\end{itemize}
The line shapes for these three cases are illustrated in Fig.~\ref{fig:lineshape}.

%%%%%%%%%%%%%%%%%%%%%%%%%%%%%%%%%%%%%%%%%%
\subsection{Bound-state wavefunction}

In a quantum field theory, the Schr\"odinger wavefunction for a 2-particle bound state 
can be determined from the $2 \to 2$ transition amplitude  for its constituents \cite{GellMann:1951rw,Salpeter:1951sz}. 
We take the particles with masses $M_0$ and $M_{*0}$ to have incoming momenta $\bm q_0$ and $\bm q_1$,
outgoing momenta $\bm q_0'$ and $\bm q_1'$ with $\bm q_0'+ \bm q_1' =\bm q_0 +\bm q_1$,
and total energy $E$ relative to the scattering threshold.
The 1PI transition amplitude for a pair of particles with nonzero total momentum 
can be obtained from the 1PI transition amplitude in the CM frame in Eq.~\eqref{Amp-E}
by replacing $E$ by the Galilean-invariant combination of the total energy and the total momentum \cite{Braaten:2015tga}:
%===========
\begin{eqnarray}
\mathcal{A}(E,\bm q_0 + \bm q_1) = 
   \frac{2\pi /\mu}{- \gamma_X  + \sqrt{-2\mu[E - (\bm q_0 +\bm q_1)^2/(2M_X) + i \Gamma_{*0}/2]}}.
\label{amp1PI}
\end{eqnarray}
%===========
For simplicity of notation, we have set $M_{*0}\!+\!M_0 = M_X$.
The connected $2 \to 2$ transition amplitude is the product of 
the 1PI amplitude $\mathcal{A}$ and a nonrelativistic propagator for each of the 4 external legs.
The propagators for the incoming particles with energies $E_0$ and $E_1$ are  
$i/(E_0 - \bm q_0^2/(2M_0)+ i \epsilon)$ and $i/(E_1 - \bm q_1^2/(2M_{*0})+ i \Gamma_{*0}/2)$.
If the external  lines for one particle in the initial state and one particle in the final state are both amputated
and if those lines are put on their energy shells, the residue of the pole can be factored into 
the product of a wavefunction that depends on the incoming momenta 
and a wavefunction that depends on the outgoing momenta  \cite{GellMann:1951rw,Salpeter:1951sz}:
%===========
\begin{eqnarray}
 \psi(\bm q_0, \bm q_1)\, 
   \frac{1}{E -[-\gamma_X^2/(2\mu) + (\bm q_0 +\bm q_1)^2/(2M_X) - i \Gamma_{*0}/2]} \,  
   \psi(\bm q_0', \bm q_1').
\label{ampconn}
\end{eqnarray}
%===========
The wavefunction is
%===========
\begin{eqnarray}
\psi(\bm q_0, \bm q_1) = 
   \frac{\sqrt{8\pi\gamma_X}}{\gamma_X^2  + \mu^2(\bm q_0/M_0 - \bm q_1/M_{*0} )^2}.
\label{psiX-q1q2}
\end{eqnarray}
%===========
This wavefunction is a function of the relative velocity of the constituents.
Setting $\bm q_0 = \bm q - \bm k$ and $\bm q_1 = \bm k$, we reproduce the wavefunction
in Eq.~\eqref{psiX2-q}  for  a bound state with momentum $\bm{q}$ 
and relative momentum $\bm{k}$, except that $\Gamma_{*0}$  has been set to 0.

%%%%%%%%%%%%%%%%%%%%%%%%%%%%%%%%%%%%%%%%%%
\subsection{Resonance Feature from the Triangle Singularity}

In our calculation of the loop amplitude $F(W)$ from the triangle singularity in Section~\ref{sec:Triangle},
we assumed that the $X$ is a narrow bound state with the sharp rest energy $E_X$.
This allowed the coupling of the $X$ to a pair of charm mesons to be described by the momentum-independent 
vertex in Eq.~\eqref{vertexX}.
In the calculation of the cross section, 
the integral over the phase space includes an integral over a delta function at the sharp energy of the bound state.
If the $X$ is not a narrow bound state, the calculation of the cross section near the peak from the triangle singularity 
is more complicated.
Whether or not the $X$ is  a narrow bound state, 
the  distribution of the rest energy for the  resonance feature 
is  given by the factor of $\mathrm{Im} [\mathcal{A}(E)]$ in Eq.~\eqref{rate-optical}.
If $X$ is a narrow bound state,
the cross section for  $e^+ e^-$ annihilation into $X+\gamma$ near the $D^{*0} \bar D^{*0}$ threshold
is given in Eq.~\eqref{sigmagammaXana}.
It has a factor of $|F(W)|^2$, where $F(W)$ is the loop amplitude in Eq.~\eqref{Fanalytic}
for a bound state  with the sharp rest energy $E_X$. 
If $X$  is not a narrow bound state,
we should allow for the additional dependence on the energy $E$ in the loop amplitude $F(W,E)$.
We should also replace the delta function in the rest energy of the bound state
by the energy distribution proportional to $\mathrm{Im} [\mathcal{A}(E)]$ in Eq.~\eqref{rate-optical}.
The factor $|F(W)|^2$  in Eq.~\eqref{sigmagammaXana} can be replaced by
%===========
\begin{equation}
\big|F(W) \big|^2 \longrightarrow 
\frac{\mu^2}{2\pi^2 |\gamma_X|} \int \!\!dE \,  \mathrm{Im} \big[\mathcal{A}(E) \big] \, \big|F(W,E) \big|^2,
\label{sub-sig}
\end{equation}
%===========
where $\mathrm{Im} [\mathcal{A}(E)]$ is given in Eq.~\eqref{ImAmp-E} and
$F(W,E)$ is given by the analytic expression in Eq.~\eqref{Fanalytic}
with the coefficients $a$, $b$, and $c$ replaced by 
%============
\begin{subequations}
\begin{eqnarray}
a &=&   k^2+i M_{*0}\Gamma_{*0} ,
    \\
b(E) &=&  -\big[   (\mu/M_0)^2   q^2 +   k^2 - 2 \mu E \big] - i  (\mu/M_0)M_{*0}\Gamma_{*0} ,
\label{coeffb-E}
    \\
c(E) &=&  (\mu/M_0)^2  q^2 .
\end{eqnarray}
\label{coefficientsabc-E}%
\end{subequations}
%===========
The factor of $1/|\gamma_X|$ in Eq.~\eqref{sub-sig} cancels the factor of $|\sqrt{\gamma_X}\, |^2$ from $|F(W,E)|^2$,
so the only dependence on $\gamma_X$ comes from $\mathrm{Im} [\mathcal{A}(E)]$.
The variable $q$ that appears in the coefficients $b$ and $c$ is the function of $W-E$ that satisfies
%==============
\begin{equation}
W = (q -\delta)+ \frac{q^2}{2M_X} +E.
\label{Econservation-E}
\end{equation}
%==============
The variable $q$ appears as a multiplicative factor in the analytic expression for $F(W)$  in Eq.~\eqref{Fanalytic}.
It also appears in the factor multiplying  $|F(W)|^2$ in the cross section in Eq.~\eqref{sigmagammaXana}.
For those terms $q$, we can use the solution to Eq.~\eqref{Econservation-E} with $E=0$.
The dependence of $q$ on $E$ is only essential in the argument of the logarithm in Eq.~\eqref{Fanalytic}.

%%%%%%%%%%%%%%%%%%%%%%%%%%%%%%%%%%%%%%%%%%%%%%%%
\begin{figure}[t]
\includegraphics*[width=0.8\linewidth]{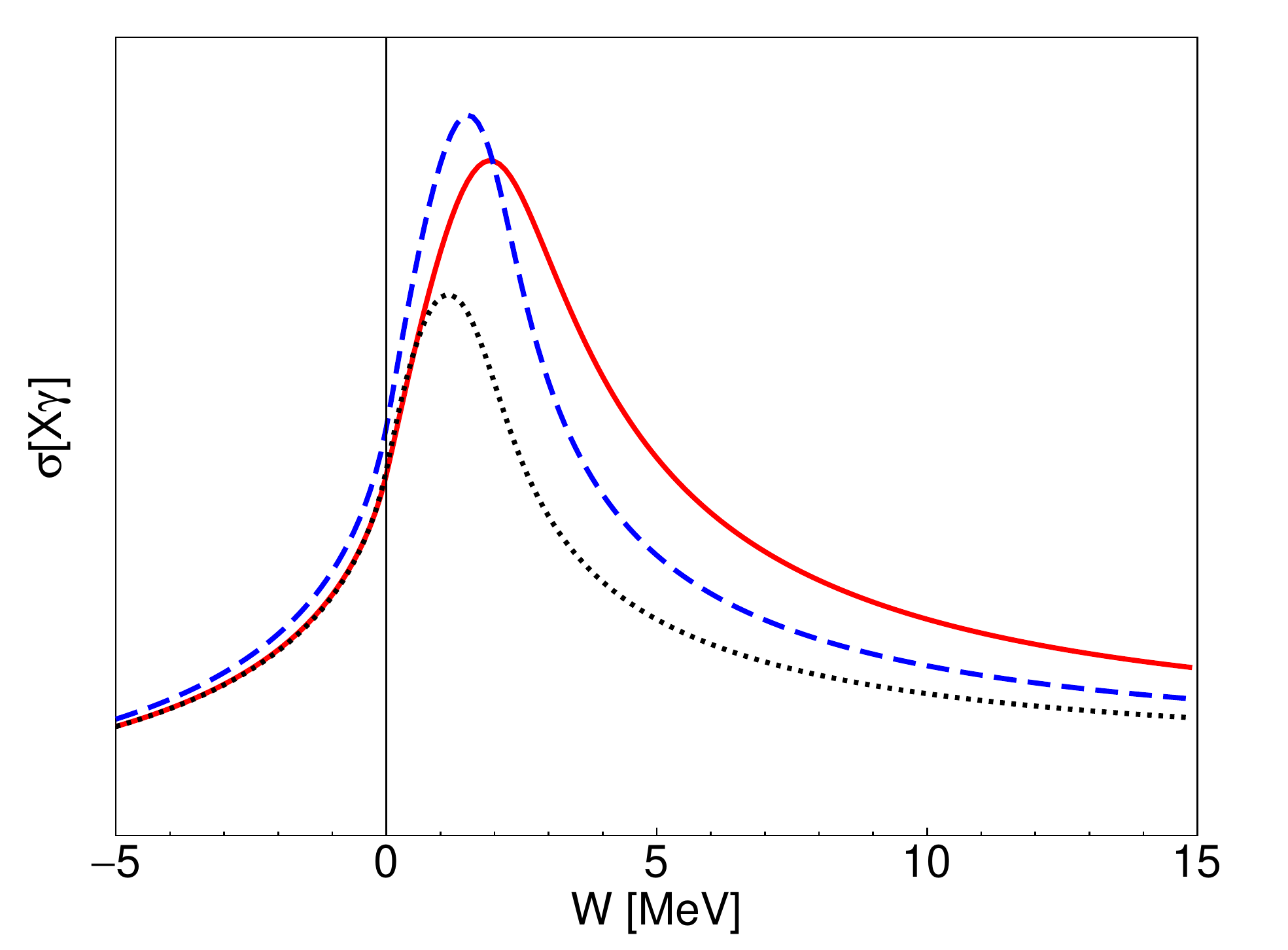} 
\caption{
Cross section for $e^+ e^- \to X \gamma$ as a function of the 
center-of-mass energy $W$ relative to the $D^{*0} \bar D^{*0}$ threshold.
The curves are for a narrow bound state with $E_X = -0.30$~MeV (solid red), 
a zero-energy resonance with $E_X =0$ (dashed blue), 
and a virtual state with $E_X = +0.30$~MeV (dotted black).
The normalizations of the cross sections are arbitrary, but their relative normalizations 
are determined by the line shapes in Fig.~\ref{fig:lineshape}.
}
\label{fig:sigmatri}
\end{figure}
%%%%%%%%%%%%%%%%%%%%%%%%%%%%%%%%%%%%%%%%%%%%%%

In Fig.~\ref{fig:sigmatri}, we compare the cross section for $X\gamma$ near the peak 
from the triangle singularity for the three resonance cases itemized in Section~\ref{sec:SDproduction}.
The corresponding line shapes for the three resonance cases are illustrated in Fig.~\ref{fig:lineshape}.
The relative normalizations for the cross sections for the narrow bound state, the zero-energy resonance, 
and the virtual state are all determined
by the condition that the three line shapes in Fig.~\ref{fig:lineshape} have the same area.
The cross section for the narrow bound state has a shape very similar to that for the bound state 
with  $|E_X| =0.30$~MeV  in Fig.~\ref{fig:sigma-bindingE}.
The height of the peak in the cross section is a little larger for the zero-energy resonance 
than for the narrow bound state, and it is smaller for the virtual state.  The peak for
the zero-energy resonance and the virtual state is at a slightly lower energy
and the full width at half maximum is smaller.
However the three cross sections in Fig.~\ref{fig:sigmatri} have roughly the same shape.
Thus the peak from the triangle singularity cannot easily discriminate between
a narrow bound state and other possibilities for the character of the resonance.

%\newpage

%%%%%%%%%%%%%%%%%%%%%%%%%%%%%%%%%%%%%%%%%%

 \end{document}